\documentclass[aps,prd,preprint,eqsecnum,nofootinbib]{revtex4}
\usepackage{epsfig}
\usepackage{amssymb,amsfonts,amsmath,bbold}
 
\def\I{{\bf I}}

\def\K{{\mathcal K}}
\def\G{{\mathcal G}}
\def\M{{\mathcal M}}
\def\bzero{\beta_0}

\def\CA{C_A}

\def\as{\ensuremath{\alpha_{s}}}
\def\a0{\alpha_0}

\def\vep{\varepsilon}
\def\f{{\rm f}}

\def\MSbar{$\overline{{\rm MS}}$}

\def \ep{\epsilon}

\def\bea {\begin{eqnarray}}
\def\eea {\end{eqnarray}}

\def\be {\begin{equation}}
\def\ee {\end{equation}}
 
\begin{document}
 
 \begin{flushright}
 YITP-SB-06-28\\
 SLAC--PUB--11969\\
\today
\end{flushright}
 
 \begin{center}
 
{\Large \bf The Two-loop Soft Anomalous Dimension Matrix\\
\medskip
and Resummation at Next-to-next-to Leading Pole }

\bigskip

{\bf \large S.\ Mert Aybat$^1$, Lance J. Dixon$^2$, George Sterman$^1$}

\bigskip

{\it
$^{1}$ C. N. Yang Institute for Theoretical Physics, Stony Brook
  University, SUNY,\\
  Stony Brook, NY 11794-3840, USA \\
$^{2}$ Stanford Linear Accelerator Center, Stanford University,\\
  Stanford, CA 94309, USA }

\end{center}

\bigskip

\begin{abstract}
  We extend the resummation of dimensionally-regulated amplitudes to
next-to-next-to-leading poles. This requires the calculation of two-loop
anomalous dimension matrices for color mixing through soft gluon exchange. 
Remarkably, we find that they are proportional to the corresponding one-loop matrices. Using the color generator notation, we reproduce the two-loop single-pole quantities ${\bf H}^{(2)}$ 
introduced by Catani for quark and gluon elastic scattering.  Our results also make
possible threshold and a variety of other resummations at next-to-next-to
leading logarithm. All of these considerations apply to $2\to n$  processes
with massless external lines.
\end{abstract}

\maketitle

\section{Introduction}

The description of partonic hard scattering in quantum chromodynamics
(QCD) is central to the analysis of final states at hadronic colliders.
The calculation of  cross sections for such processes
requires a combination of virtual and real radiative corrections,
organized according to underlying factorization theorems.
This is the case for higher-order calculations to next- or 
next-to-next-to-leading order in $\alpha_s$ (NLO, NNLO, \dots). 
It holds as well as for resummed cross sections, in which
selected corrections associated
with soft and collinear gluon radiation are organized, at leading, next-to-leading
or next-to-next-to-leading logarithms (LL, NLL, NNLL, \dots )
to all orders in $\as$.     

In both fixed-order and resummed 
calculations the coherence properties of soft gluon radiation
play an essential role.  An anomalous dimension
matrix for inclusive wide-angle soft gluon radiation was introduced 
in Refs.~\cite{Sen83,botts89},
and computed to leading order for quark and gluon scattering processes 
in Ref.~\cite{KOS}.
The one-loop matrix of soft anomalous dimensions has been applied to 
the NLL threshold resummation of jet 
cross sections~\cite{KOSjet,KOjet} 
and of distributions of event-shape variables~\cite{BSZ,Boncianietal} 
that are ``global" in the sense of Ref.~\cite{DS01}.
At two loops, the same matrix, combined with resummed form factors,
was shown in Ref.~\cite{TYS} to control the single
infrared poles of dimensionally-regularized partonic scattering
amplitudes in $\vep=2-D/2$.  In this paper we will show how to compute this
matrix directly at two loops, from a relatively limited set of 
diagrams in the eikonal approximation, using Wilson lines,
giving as an explicit example quark-antiquark scattering.   

The full analysis given below applies to any $2\to n$ partonic amplitude
in dimensional regularization.  The two-loop soft anomalous dimension
matrix allows the exponentiation of next-to-next-to-leading infrared poles,
which appear in the combination $\as^n(1/\vep)^{n-1}$ in the exponent,
a level equivalent to next-to-next-to-leading logarithms.
The resulting resummed amplitudes can be expanded out to the 
two-loop order, and the poles in $\vep$ can be compared to explicit
two-loop scattering amplitudes, for example the basic $2\to2$
scattering processes~\cite{Twoloopqqqqqqgg,Twoloopgggg,Twoloopgggghel}.
Those poles were expressed in terms of the color-space
notation~\cite{catani96}
and the organization
of two-loop singular terms presented in 
Ref.~\cite{catani98}.  (Related work at one loop was performed 
in Refs.~\cite{gieleglover92,kst94}.)
We will verify that the expansion of the
resummed amplitudes to two loops matches precisely the
full infrared pole structure of the known two-loop scattering
amplitudes, including the single poles in $\varepsilon$.
Remarkably, we will find, as reported in Ref.~\cite{adsletter06}, 
that the two-loop anomalous dimension matrix is related to the 
one-loop matrix by a constant, the same constant, $K$, appearing in
the DGLAP splitting kernel, that relates the one and two-loop 
anomalous dimensions for the Sudakov form factor.  
(The analogous matrix appears in the electroweak Sudakov corrections
to four-fermion scattering, and has been extracted at two loops from the 
QCD four-quark scattering amplitude in Ref.~\cite{JKPS}.)
The simplicity of this result
will facilitate the development of practical resummed
cross sections with color exchange at NNLL.

This paper is organized as follows.
The next section reviews the collinear and infrared factorization 
of exclusive amplitudes. In that section, we provide a new explicit 
scale-setting choice for the soft function, which is necessary to 
define the scales of logarithms in the relevant anomalous dimensions.   
The third section describes the expansion of the jet functions to 
two loops.  Here we describe a new ``minimal" reorganization of the 
factorized amplitude, to facilitate the comparison to fixed-order calculations.
In the fourth section, we describe in detail the one- and two-loop 
calculations necessary to determine the soft anomalous dimension matrix, 
for the specific case of quark-antiquark scattering.  
Here, we will employ the eikonal approximation, and the scale-setting 
choice for the soft function from Sec.~2.  
We show that diagrams attaching gluons to three different eikonal
lines either vanish, or represent the exponentiation of the one-loop
soft matrix.  We close Sec.~4 by generalizing these
calculations to arbitrary flavors for incoming partons and arbitrary
flavors and numbers of outgoing partons.  To do so, we present
the color-mixing anomalous dimension matrix in the color-space notation
of Ref.~\cite{catani96,catani98}.   Finally, in Sec.~5 we employ this notation, 
along with results of Sec.~4 for the soft anomalous dimensions and 
known two-loop elastic form factors for quarks and gluons, 
to give the explicit form of the two-loop single pole terms in $\vep$,
for arbitrary $2\to n$ partonic processes in QCD.
We show that these pole terms agree with the single-pole 
``${\bf H}^{(2)}$" terms found in NNLO $2\rightarrow 2$
calculations~\cite{Twoloopqqqqqqgg,Twoloopgggg,Twoloopgggghel,ee3partonNNLO}
whose poles have been organized according to the formalism 
of Ref.~\cite{catani98}.
Our results also agree with the proposal of 
Ref.~\cite{BDK04} for the single poles for the
case of $2\to n$ gluon processes, which was based on the consistency
of collinear factorization of amplitudes.
We provide an appendix with explicit forms of Sudakov anomalous dimensions,
and two appendices illustrating calculations of soft anomalous
dimensions using eikonal methods.  The final appendix details the 
computation of a particular commutator of color-space matrices, which is 
needed to compare our results with the explicit NNLO calculations.

\section{Factorized amplitudes in dimensional regularization}
\label{FactorizedAmpSection}

Our considerations apply to $2\to n$ scattering processes,
denoted as ``\,f\,",
\begin{equation}
\label{partproc}
\f : \quad f_1(p_1,r_1) + f_2(p_2,r_2) 
\to f_3(p_3,r_3) + f_4(p_4,r_4) + \dots + f_{n+2}(p_{n+2},r_{n+2})\,.
\end{equation}
The labels $f_i$ refer to the flavor of the participating partons, 
each of momenta $\{ p_i \}$
and color $\{r_i\}$. The amplitude for this process, ${\M}^{[\f]}$, 
is a color tensor with indices associated with the external 
partons $\{r_i\} = \{ r_1, r_2, \dots \}$. It is convenient to express 
these amplitudes in a basis of $C$ independent color tensors,
$\left(c_I\right)_{\{r_i\}}$, so that~\cite{KOS,catani98}
\begin{eqnarray}
\label{amp}
\M^{[\f]}_{\{r_i\}}\left(\beta_j,\frac{Q^2}{\mu^2},\as(\mu^2),\vep \right)
&=&
\sum_{L=1}^C \M^{[\f]}_{L}\left(\beta_j,\frac{Q^2}{\mu^2},\as(\mu^2),\vep \right)
\, \left(c_L\right)_{\{r_i\}}
\nonumber\\
&=& \left| {\cal M}_\f\right \rangle\, ,
\end{eqnarray}
where the ket may be thought of as a vector $\M^{[\f]}_L$
with $C$ elements in the space of color tensors $c_I$.  
We will analyze these amplitudes at fixed momenta $p_i$ 
for the participating partons, which we represent as
\bea
p_i = Q\beta_i \, , \quad \beta_i^2=0\, ,
\label{betaidef}
\eea
where the $\beta_i$ are four-velocities, and where
$Q$ is an overall momentum scale.  For the
purposes of this analysis, and to compare
with existing NNLO calculations, we take all of the partons
massless, as indicated.   To be specific,  we may take $\beta_1\cdot \beta_2 =1$
for the incoming partons in Eq.~(\ref{partproc}), which
implies $Q^2=s/2$, but this is not necessary.

In dimensional regularization ($D=4-2\vep$), on-shell amplitudes may be
factorized into jet, soft and hard functions, which describe the
dynamics of partons collinear with the external lines, soft exchanges
between those partons, and the short-distance scattering process,
respectively.  This factorization follows from the general
space-time structure of long-distance contributions
to elastic processes~\cite{Akhoury}.  A formal proof for the case $n=2$
in QCD 
(quark-quark scattering) was presented long ago~\cite{Sen83}.  

The general form of the factorized amplitude is 
\begin{eqnarray}
\label{facamp}
\M^{[\f]}_{L}\left(\beta_i,\frac{Q^2}{\mu^2},\as(\mu^2),\vep \right)
&=&J^{[\f]}\left(\frac{Q'{}^2}{\mu^2},\as(\mu^2),\vep \right)
S^{[\f]}_{LI}\left(\beta_i,\frac{Q'{}^2}{\mu^2},\frac{Q'{}^2}{Q^2},\as(\mu^2),
                     \vep \right) 
\nonumber\\
&\ & \hspace{0.2mm} 
\times \ H^{[\f]}_{I}\left(\beta_i,\frac{Q^2}{\mu^2},\frac{Q'{}^2}{Q^2},\as(\mu^2)
\right)\, ,
\end{eqnarray}
where $\mu$ is the renormalization scale.
 $J^{[\f]}$ is the product of jet functions
for each of the external partons, as above denoted
collectively by [f], $S^{[\f]}$ is the soft function,
and $H^{[\f]}$ is the short-distance function.
For example, when the process is
$1 + 2 \to 3 + 4$, the product of jet functions is
\begin{eqnarray}
\label{psidesc}
J^{[\f]}\left(\frac{Q'{}^2}{\mu^2},\as(\mu^2),\vep \right)
&\equiv &
\prod_{i=1,2,3,4}J^{[i]}\left(\frac{Q'{}^2}{\mu^2},\as(\mu^2),\vep
\right) \, .
\end{eqnarray}

Construction of the soft and jet functions requires the specification
of at least one independent momentum scale, $Q'$, which plays
the role of a factorization scale.  Such a scale, 
distinct from $Q$ and $\mu$, may be useful
when one or more invariants obey strong ordering.  Here,
however, we shall consider ``fixed-angle" scattering configurations,
in which the parameter $Q$ sets the scale for all invariants,
up to numbers of order unity.   With this in mind,  we will
simplify Eq.~(\ref{facamp}) somewhat, and pick  $Q'=\mu$,
that is, equal factorization and renormalization scales.
Both the soft and jet functions then depend on $\as(\mu^2)$
only, and we will suppress their $Q'{}^2$ dependence,
now expressing the same amplitude as
\begin{eqnarray}
\label{simfacamp}
\M^{[\f]}_{L}\left(\beta_i,\frac{Q^2}{\mu^2},\as(\mu^2),\vep \right)
&=&J^{[\f]}\left(\as(\mu^2),\vep \right)
S^{[\f]}_{LI}\left(\beta_i,\frac{Q^2}{\mu^2},\as(\mu^2),\vep \right) \nonumber
\\
&&\times ~~H^{[\f]}_{I}\left(\beta_i,\frac{Q^2}{\mu^2},\as(\mu^2)
\right)\, ,
\end{eqnarray}
that is, we suppress dependence on  those variables that are
set to unity by our choice of scales.  

Clearly, any jet-soft-hard factorization
of the sort described above
 is unique only up to
finite factors in the various functions.  There is
an additional ambiguity between the
jet and soft functions at the level of a single infrared
pole 
per loop in dimensional regularization.  
In the remainder of this section, we will
provide specific definitions for the jet
and soft functions that will enable us to define
and resum them unambiguously, and which will be useful in
our calculations below.  We begin with the jet functions.

\subsection{The jet functions and the Sudakov form factor}

The factorization~(\ref{facamp}) holds for any exclusive
amplitude, including the elastic, or Sudakov, form factor.
A very natural definition of the jet functions is, therefore,
the square root of the form factor~\cite{TYS}.
Here, we will choose the case of the elastic scattering
form factor with a color-singlet source, and spacelike momentum transfer.
Reverting to the general case of jet momentum scale $Q'{}^2$, not
necessarily equal to the  renormalization scale, this is
\begin{eqnarray}
\label{jsudakov}
J^{[i]}\left(\frac{Q'{}^2}{\mu^2},\as(\mu^2),\vep\right)
  = J^{[\bar i]}\left(\frac{Q'{}^2}{\mu^2},\as(\mu^2),\vep\right)
&=& \left[\M^{[i  \to i]}
  \left(\frac{Q'{}^2}{\mu^2},\as(\mu^2),\vep\right)\right]^{\frac{1}{2}} \,.
\end{eqnarray}
Below, we shall take $\mu$ as the \MSbar~renormalization scale, 
$\mu^2 = \mu_0^2 \, {\rm exp}[-\vep(\gamma_E-{\rm ln}(4\pi))]$.
With this choice,  we may rely on the explicit form of
the quark spacelike electromagnetic Sudakov form factor in $D=4-2\vep$
dimensions.  A similar definition may be given for gluon
jets in terms of matrix elements of conserved, singlet
operators.  In either case, the all-orders expression
for the (square root of the) resummed form factor, organizing all pole
terms, and implicitly specifying all finite terms
of the jet defined as in Eq.~(\ref{jsudakov}), 
is~\cite{magneags,magnea,Collins89}
\begin{eqnarray}
\label{solutionev}
J^{[i]}\left(\frac{Q'{}^2}{\mu^2},\as(\mu^2),\vep\right)
&=& 
{\rm exp}\Biggl\{
~\frac{1}{4} \int_{0}^{Q'{}^2}\frac{d\xi^2}{\xi^2}
\Biggl[
\K^{[i]}(\as(\mu^2),\vep)
\nonumber \\
&&\hskip34mm
+\,\G^{[i]}
\left(-1,\bar\as\left(\frac{\mu^2}{\xi^2},\as(\mu^2),\vep\right),\vep\right)
\nonumber \\
&&\hskip34mm
+\,\frac{1}{2}\int_{\xi^2}^{\mu^2} 
\frac{d\tilde{\mu}^2}{\tilde{\mu}^2}
\gamma^{[i]}_{K}\left(\bar\as\left(\frac{\mu^2}{\tilde{\mu}^2},\as(\mu^2),\vep
  \right)\right)
\Biggr] ~\Biggr\} \,,~~~
\end{eqnarray}
where we use a notation for the running coupling that
emphasizes its re-expansion in terms of the coupling at fixed scale $\mu$.
For our purposes below, we shall need only the ``leading" form
of the running coupling,
\begin{eqnarray}
\label{asinD}
\bar\as\left(\frac{\mu^2}{\tilde{\mu}^2},\as(\mu^2),\vep\right) &=&
\as(\mu^2)\left(\frac{\mu^2}{\tilde{\mu}^2} \right)^\vep
\sum_{n=0}^{\infty}\left[\frac{\bzero}{4\pi\vep}
\left(\left(\frac{\mu^2}{\tilde{\mu}^2} \right)^\vep-1\right) \as(\mu^2)
\right]^n \,,
\end{eqnarray}
with the one-loop coefficient
\begin{equation}
\bzero = \frac{11}{3}\CA - \frac{4}{3} T_F n_F \,.
\end{equation}
In  the expression for the jet functions above,
the choice $Q'{}^2=\mu^2$ can be imposed trivially.
The functions $\K^{[i]}$, $\G^{[i]}$ and $\gamma^{[i]}_K$ 
are anomalous dimensions that can be determined 
by comparison to fixed-order calculations of the Sudakov form factors
for quarks and gluons.  These form factors are now known in QCD
up to three loops~\cite{EMformfactor,Harlander00,MVVformfactor}. 
Notice that the coupling in the
argument of $\K^{[i]}$ is fixed at $\mu$, so that
the integral of this term alone is not well-defined
at $\xi^2=0$ even for $D\ne 4$.  This apparent divergence, however,
is cancelled by contributions from the upper limit of
the $\tilde{\mu}^2$ integral of the anomalous dimension
$\gamma^{[i]}_K$, and relates the latter to $\K^{[i]}$
order-by-order in perturbation theory.  We will provide
explicit expansions for these functions in Appendix~\ref{AnomDimAppendix}.

\subsection{The soft function}

We will broadly follow Ref.~\cite{KOS} in
the definition of the soft function for partonic
amplitudes, although we will modify certain details
in the construction.
The fundamental observation of Ref.~\cite{KOS} is that the
soft function, describing color exchange between
the jets, is independent of collinear dynamics, and
may be constructed from an eikonal amplitude,
that is, the vacuum expectation of products
of ordered exponentials.  For each external parton of
flavor $f_i$, we introduce a nonabelian path-ordered phase operator,
\bea
\Phi^{[f_i]}_{v_i}(\sigma',\sigma) = 
P \exp\, \left[\, -ig \int_\sigma^{\sigma'} d\lambda\;  v_i \cdot A^{[f_i]}(\lambda v_i)\, \right]\, ,
\label{Phidef}
\eea
where $v_i^\mu \sim \beta_i^\mu$ is a four-velocity.   
For specific calculations at 
two loops, it will be useful to choose these velocities to
be slightly timelike, 
\bea
\label{vlength}
0< v_i^2  \ll 1 \, .
\eea  
The ``opposite moving"
velocity $\bar v_i^\mu$ projects out the large component
of $v_i^\mu$.  The gauge field $A^{[f_i]}$ is a matrix in
the representation of parton $i$.
In the  construction of the
soft function, we will eventually take all $v_i^2\to 0$, or
equivalently,  $v_i^\mu \to \beta_i^\mu$.
In perturbation theory, the operators $\Phi^{[f_i]}_{v_i}(\infty,0)$
and  $\Phi^{[f_i]}_{v_i}(0,-\infty)$ respectively
generate outgoing and incoming eikonal lines in the $v_i$-directions.
The eikonal sources couple to gluons at vertices in the color representation of parton $i$.
An essential feature of these diagrams is that
they are invariant under rescalings of the velocities, $v_i \to  \sigma v_i$.

We are now ready to construct eikonal multi-point amplitudes from
products of ordered exponentials, tied together
by the same color tensors, $c_L$ that appear in the expansion
of the partonic amplitudes, Eq.~(\ref{facamp}).  For the 
$2\to 2$ case, $1+2 \to 3+4$, this gives
\bea
W^{[\f]}_I{}_{\{r_k\}}
= (c_L)_{\{r_k\}} \,
W_{LI}^{[\f]}
\Bigl({\textstyle{\frac{v_i \cdot v_j}{\sqrt{v_i^2v_j^2}}}}\Bigr)
&=&
\sum_{\{d_i\}}
\langle0|\, \Phi_{v_4}^{[f_4]}(\infty,0)_{r_4,d_4}\; 
\Phi_{v_3}^{[f_3]}(\infty,0)_{r_3,d_3}\cr
&\ & \hspace{-15mm} \times
\left( c_I\right)_{d_4d_3,d_2d_1}\; 
\Phi_{v_1}^{[f_1]}(0,-\infty)_{d_1,r_1}
\Phi_{v_2}^{[f_2]}(0,-\infty)_{d_2,r_2}\, |0\rangle \,.~~~~~
\label{eq:wivertex}
\eea
Such a product is gauge invariant.
The eikonal amplitude, or web function, 
$W$ depends in general on both the invariants $v_i\cdot v_j$
and the invariant lengths $v_i^2$.
The basic observation of Ref.~\cite{KOS} is that all potentially
collinear divergent ratios factorize from dependence on
wide-angle  radiation for eikonal as well
as partonic amplitudes.  We can use this factorization
to isolate the soft function  systematically, using only
calculations in the eikonal approximation.

Because of the factorization of collinear singularities,
such dependence is universal, depending only on the number and 
flavors of the external jets.   In particular,
as observed above, form factors, with two external lines
and trivial color flow, generate the same collinear dependence.
Thus,  all collinear dependence cancels in the ratio of our four-point eikonal
amplitude $W_I$ and the product of two eikonal form factors,
just as in the ratio of the four-point partonic amplitudes
to the corresponding form factors.  We shall define $S_{LI}$
by this ratio.   Notice that information on color flow
is not affected at all by the eikonal jet functions, which like
partonic jets, are diagonal in color.  
Thus, we define
\bea
S_{LI}^{[\f]} \left( \frac{\beta_i\cdot \beta_j}{u_0} \right)= 
\lim_{v^2\to 0} \frac{W_{LI}^{[\f]} \left ( \frac{v_i\cdot v_j}{v^2} \right)}
{\prod_{i\in\f} \left [ \,  
W^{(i\rightarrow i)} \left (\frac{u_0}{v^2} \right )\right ]^{1/2}} \, ,
\label{Sdef} 
\eea
where as above the velocities $\beta_i$ are the lightlike limits of the $v_i$.
The denominators are eikonal versions of the elastic form
factors, defined with incoming velocities $v_i$ and outgoing $\bar v_i$,
where $v_i^2 = \bar v_i^2=v^2$ and $v_i\cdot \bar v_i = u_0$, with $u_0$
a constant of order unity, independent of $i$, 
namely
\bea
\label{Wiidef}
W^{(i\rightarrow i)} \left( \frac{u_0}{v^2} \right) =
\langle 0|\, \Phi^{[f_i]}_{\bar v_i}(\infty,0)
           \,\Phi^{[f_i]}_{v_i}(0,-\infty)\, |0\rangle \, .
\eea
This form factor generates the square of the collinear poles 
associated with the eikonal jet of flavor $i$ in $W_{LI}$, and hence
the soft function~(\ref{Sdef}) is free of collinear divergences.
We may thus take the lightlike limit for the velocities 
to define the soft function in the ratio.  

Equation~(\ref{Sdef}) allows us to compute the soft function,
once we determine how to choose the variable $u_0$,
so that we may match the eikonal calculation to the 
partonic amplitude.  We can determine the correct
choice as follows.

We first re-express Eq.~(\ref{simfacamp}) for
the partonic amplitude, converting it into an expression for
the soft function as a ratio analogous to Eq.~(\ref{Sdef}),
\begin{eqnarray}
\label{solveSpart}
S^{[\f]}_{LI}\left(\beta_i,\frac{Q^2}{\mu^2},\as(\mu^2),\vep \right)
H^{[\f]}_{I}\left(\beta_i,\frac{Q^2}{\mu^2},\as(\mu^2)
\right)=
\frac{
\M^{[\f]}_{L}\left(\beta_i,\frac{Q^2}{\mu^2},\as(\mu^2),\vep \right)}
{J^{[\f]}\left(\as(\mu^2),\vep \right)}\, .
\end{eqnarray}
This simple result enables us to set the scale $u_0$ in the definition 
of the eikonal form factors of Eq.~(\ref{Sdef}).  In Eq.~(\ref{solveSpart}), 
$S$ can depend on
the velocities only through the ratios $\beta_i\cdot\beta_jQ^2/\mu^2$.
When $S$ is calculated in this way from the ratio
of partonic quantities, $Q$ sets the scale of all momenta in the 
amplitude, and $\mu$,
the factorization scale in Eq.~(\ref{simfacamp}), may be reinterpreted
as the momentum transfer in the form  factors that
define the jet functions.  When calculated from the eikonal
ratio, on the other hand, $S$ depends only on the variables
$\beta_i\cdot \beta_j/u_0$.  To match the soft function computed
in eikonal approximation with the partonic amplitude, we
need only require
\bea
\frac{\beta_i\cdot \beta_j}{u_0} = \frac{Q^2\beta_i\cdot \beta_j}{\mu^2}
\qquad \rightarrow \qquad
u_0 = {\mu^2\over Q^2}\, .
\label{u0def}
\eea
This relation will be used in our explicit calculations later.
We are now ready to provide an all-orders expression 
for the soft function, analogous to Eq.~(\ref{solutionev})
for the jet functions.

\subsection{Resumming the soft function}

We will use the $\overline{\rm MS}$ scheme for 
renormalization throughout.
Before renormalization, all of the purely 
eikonal amplitudes discussed in the
previous subsection give (only)  scaleless integrals
in perturbation theory.  Such integrals
vanish identically in dimensional regularization.
In fact, these functions are only nontrivial because of
renormalization, with every infrared pole resulting
from the subtraction of a corresponding  ultraviolet pole.
This is the case whether or not $W$ is collinear-regulated
by introducing masses for its eikonal phases. 

Thus, for both the web function $W$ and the soft function $S$, we have 
(suppressing indices)
\begin{eqnarray}
\label{Eq:renS}
W^{[\f]}_{\rm bare} &=& 1\, =\, Z_{W_\f}(\as(\mu),\vep)\,W^{[\f]}_{\rm ren}\, ,
\nonumber\\
S^{[\f]}_{\rm bare} &=& 1\, =\, Z_{S_\f}(\as(\mu),\vep)\,S^{[\f]}_{\rm ren}\, ,
\end{eqnarray}
and similarly for the eikonal form factors in the ratio~(\ref{Sdef}).
Both $S$ and $W$ are therefore defined entirely
by their anomalous dimension matrices,
\bea
\left(\Gamma_A\right)_{IJ} &=& \left(Z_A^{-1}\right)_{IK}\, \frac{d \left(Z_A\right)_{KJ}}{d\ln\mu}
\nonumber\\
&=& \left(Z_A^{-1}\right)_{IK}\, \beta(g,\vep)\, \frac{\partial \left(Z_A\right)_{KJ}}{\partial g}\, ,
\label{Gammaadef}
\eea
which are given in any minimal scheme by the residues 
$Z_{A,1}^{(k)}$, with $A=W_\f$ or $S_\f$,
of single ultraviolet poles in $1/\vep$, 
at $k$th order in the expansion
\bea
Z_A=1+\sum_{k=1}^{\infty}\,\left(\frac{\alpha_s}{\pi}\right)^k
\, Z_A^{(k)}(\vep)
= \sum_{k=1}^{\infty}\,\left(\frac{\alpha_s}{\pi}\right)^k\,  
\sum_{n=1}^k\,Z_{A,n}^{(k)}\,\left(\frac{1}{\vep}\right)^n\, .
\eea
Then, for example, from the one-loop bare integrals we 
find the one-loop anomalous dimension from the residues of
the one-loop single ultraviolet poles,
\begin{equation}
\label{Eq:gamma_oneloop}
\Gamma_A^{(1)} = - 2\,Z_{A,1}^{(1)}\, ,
\end{equation}
where $\Gamma_A^{(n)}$ is the $n$th order coefficient of
$(\as/\pi)^n$ in $\Gamma_A$.
Similarly, to order $\mathcal{O}(\alpha_s^2)$, after one-loop
renormalization we find the two-loop anomalous dimensions
from the two-loop single poles,
\begin{equation}
\label{Eq:gamma_twoloops}
\Gamma_A^{(2)} = - 4\,Z_{A,1}^{(2)}\, .
\end{equation}
{}From the definition of $S$, Eq.~(\ref{Sdef}),
the soft anomalous dimension matrix is found from the
matrix for the corresponding eikonal amplitude by simply
subtracting the anomalous dimensions for the 
eikonal jets.  We denote the latter by
$\Gamma_2^{[i]}(u_0/v^2,\as)$, and write
\bea
\Gamma_{S_\f,IJ} \left({\beta_i\cdot\beta_j \over u_0},\as \right) &=&
\lim_{v^2\to 0}\, \Biggl[
\Gamma_{W_\f,IJ} \biggl({v_i \cdot v_j\over v^2},\alpha_s\biggr)
- \delta_{IJ}\; \sum_{i\in\f} 
\Gamma_2^{[i]}\biggl({u_0\over v^2}, \alpha_s \biggr) \Biggr]\, .
\label{GammaSIJdef}
\eea
In $\Gamma_{S_\f}$, all sensitivity to collinear dynamics,
and therefore to the choice of $v^2$, is cancelled,
and the coefficients depend only on the invariants
$\beta_i\cdot \beta_j$.

The matrix renormalization group equation for the 
eikonal amplitude $S_{IK}^{[\f]}$ is then
\bea
\biggl( \mu {\partial\over \partial\mu}
  + \beta(g,\vep) {\partial\over \partial g} \biggr)
 S_{IK}^{[\f]} &=& 
-\ \Gamma_{S_\f,IJ}
\left({ \beta_i\cdot\beta_jQ^2\over \mu^2 }, \alpha_s\right)\; S_{JK}^{[\f]}
\,,
\label{Gammawijdef}
\eea
from which we can solve directly for $S$ as a path-ordered exponential,
\begin{equation}
\label{expoS}
{\bf S}_\f \left(\frac{\beta_i\cdot \beta_j}{u_0},\as(\mu^2),\vep \right)
\,=\,
{\rm P}~{\rm exp}\left[
\, -\; \frac{1}{2}\int_{0}^{\mu^2} \frac{d\tilde{\mu}^2}{\tilde{\mu}^2}
{\bf \Gamma}_{S_\f} \left({\beta_i\cdot\beta_j \over u_0} ,\bar\as\left(\frac{\mu^2}{\tilde{\mu}^2},\as(\mu^2),\vep\right)\right) \right]\, ,
\end{equation}
where boldface (with a subscript for flavor flow) indicates a matrix.
In summary, the matrix of anomalous dimensions, and hence the soft matrix itself,
can be computed order-by-order purely from eikonal diagrams.

\section{The jet functions to two loops}

In this section, we expand the jet functions in  the
factorized amplitude~(\ref{simfacamp}) to fixed (second)
order in $\as$, in a form that is convenient for
comparison to explicit partonic calculations.

To determine the jet anomalous dimensions, as well as
to use the resummed forms of the jet and soft functions
with fixed-order calculations, we re-expand the
running coupling in terms of a coupling at fixed scale.  It is important
to do so consistently in dimensional regularization, using the
explicit form for the running coupling, Eq.~(\ref{asinD}).  
It will also be convenient to 
use Eq.~(\ref{solutionev}) as a starting-point to isolate the truly universal
pole terms in the logarithm of the jet function,
separating them from the finite terms.   
To this end, we introduce the notation
\bea
\ln J^{[i]}(\as(\mu^2),\vep)
&=& 
 \sum_{n=1}^\infty \left( \frac{\as(\mu^2)}{\pi} \right)^n\, 
\sum_{m=1}^{n+1}\, \frac{E_m^{[i] \, (n)}(\vep)}{\vep^m}
+
\sum_{n=1}^\infty \left( \frac{\as(\mu^2)}{\pi} \right)^n\, 
e^{[i] \, (n)}(\vep)
\nonumber\\
&=&
 E^{[i]}(\as(\mu^2),\vep)\ +\ e^{[i]}(\as(\mu^2),\vep)\,,
\label{lnJfact}
\eea
in terms of the coupling $\as(\mu^2)$ at fixed scale $\mu$.  As in
Eq.~(\ref{simfacamp}), we set the jet factorization scale $Q'=\mu$.
The pure pole terms in Eq.~(\ref{lnJfact}) have been expanded at each order as
\bea
E^{[i]\,(n)}(\vep) \equiv 
\sum_{m=1}^{n+1}\, \frac{E_m^{[i] \, (n)}(\vep)}{\vep^m}\, ,
\label{Eidef}
\eea
while the functions $e^{[i] \, (n)}(\vep)$ absorb all terms that remain finite
for $\vep=0$, order-by-order in $\as$.   The coefficients 
$E_m^{[i] \, (n)}$ and the functions $e^{[i] \, (n)}(\vep)$ are
determined, of course, by the expansions of the functions 
$\gamma_K$, ${\cal K}$
and ${\cal G}$, which depend in general on the definition~(\ref{jsudakov})
of the jet.
This separation, however,
eliminates the remaining arbitrariness in choosing the form factor 
by defining a ``minimal" jet, consisting of the exponential of pole terms only,
\bea
{\cal J}^{[i]}(\as(\mu^2),\vep) 
\equiv \exp\, \left[\, E^{[i]}(\as,\vep)\, \right]\, .
\label{calJdef}
\eea
In this notation, we rewrite our basic factorization, 
Eq.~(\ref{simfacamp}), in ``minimal" form as
\begin{eqnarray}
\label{facampmin}
\M^{[\f]}_{L}\left(\beta_i,\frac{Q^2}{\mu^2},\as(\mu^2),\vep \right)
&=&\prod_{i\in \f}{\cal J}^{[i]}\left(\as(\mu^2),\vep \right)
S^{[\f]}_{LI}\left(\frac{\beta_i\cdot \beta_jQ^2}{\mu^2},\as(\mu^2),\vep \right) \nonumber
\\
&&\times ~~{\cal H}^{[\f]}_{I}\left(\beta_i,\frac{Q^2}{\mu^2},\as(\mu^2)
\right)\, ,
\end{eqnarray}
where we have absorbed the (color-diagonal) finite factors
into the perturbative definition of the short-distance function
\bea
{\cal H}^{[\f]}_{I}\left(\beta_i,\frac{Q^2}{\mu^2},\as(\mu^2),\vep\right)
=
\exp\left[\, \sum_{i\in\f} e^{[i]}\left( \as(\mu^2),\vep \right)\, \right]\; 
H^{[\f]}_{I}\left(\beta_i,\frac{Q^2}{\mu^2},\as(\mu^2)\right)\, .
\label{calHdef}
\eea
We will also find it useful to write this expression in the color
state notation of Eq.~(\ref{amp}), as
\bea
\left|\, {\cal  M}_\f \right \rangle 
&=&\prod_{i\in \f}{\cal J}^{[i]}\left(\as(\mu^2),\vep \right)
{\bf S}_\f \left(\frac{\beta_i\cdot \beta_jQ^2}{\mu^2},\as(\mu^2),\vep \right) 
\, \left|\, {\cal H}_\f \right \rangle \, ,
\label{statecalH}
\eea
where again the matrix structure of the soft function is denoted
by boldface and where we treat ${\cal H}^{[\f]}_{I}$ 
in the notation of Eq.~(\ref {amp}).
 
Before this refactorization, the logarithm 
of the full jet function $J^{[i]}$ at two loops is given by
\bea
\ln J^{[i]}(\as(\mu^2),\vep)
&=& \frac{1}{4}\, 
 \Bigg \{  - \left( \frac{\as}{\pi} \right)\, 
 \left(\, \frac{1}{2\varepsilon^2}\, \gamma^{[i] \, (1)}_K
       + \frac{1}{\varepsilon} {\cal G}^{[i] \, (1)}(\varepsilon)\, \right)
 \nonumber\\
 &\ & \hspace{-35mm}
 +  \left( \frac{\as}{\pi} \right)^2\, 
  \left[ \frac{\beta_0}{8}\,\frac{1}{\varepsilon^2}
   \, \left( \frac{3}{4\varepsilon}\gamma_K^{[i] \, (1)} 
     + {\cal G}^{[i] \, (1)}(\varepsilon)\, \right)
 - \frac{1}{2}\left( \, \frac{\gamma_K^{[i] \, (2)}}{4\varepsilon^2} 
    + \frac{{\cal G}^{[i] \, (2)}(\varepsilon)}{\varepsilon}
 \right) \, \right] \, + \dots\, \Bigg \} \, .
\label{logJmin}
 \eea
To determine the coefficients $E_m^{[i] \, (n)}$ in the minimal two-loop
jet function, we only need to expand the functions 
${\cal G}^{[i] \, (n)}(\vep)$,
\bea
{\cal G}^{[i] \, (n)}(\vep) 
= {\cal G}^{[i] \, (n)}_0 + \vep \, {\cal G}^{[i] \, (n)}{}'(0) + \dots\, .
\label{calGnotation}
\eea
Explicit forms for these anomalous dimensions can be found in 
Appendix~\ref{AnomDimAppendix}.
In terms of these quantities, we readily find that
the full single-pole terms in the logarithm of the jet function 
are given at one and two loops by
\bea
E^{[i] \, (1)}_2 &=&  - \frac{1}{8}\, \gamma^{[i] \, (1)}_K \,,
\nonumber\\
E^{[i] \, (1)}_1 &=& - \frac{{\cal G}^{[i] \, (1)}_0}{4} \,,
\nonumber\\
E^{[i] \, (2)}_3 &=&   \frac{3\beta_0}{128}\, \gamma^{[i] \, (1)}_K \,,
\nonumber\\
E^{[i] \, (2)}_2 &=&  
 \frac{\beta_0}{32}\, {\cal G}^{[i] \, (1)}_0
- \frac{1}{32}\,\gamma^{[i] \, (2)}_K
\,, \nonumber\\
E^{[i] \, (2)}_1 
&=&
- \frac{{\cal G}^{[i] \, (2)}_0}{8} 
+ \frac{\beta_0\, {\cal G}^{[i] \, (1)}{}'(0)}{32}
\,, \nonumber\\
E^{[q] \, (2)}_1 
&=& -\, \frac{3}{8}C_F^2\, 
\left[ \frac{1}{16} - \frac{1}{2}\zeta(2) +\zeta(3)\right]
- \frac{1}{16}C_AC_F \, \left [ \frac{961}{216} +
  \frac{11}{4}\zeta(2) - \frac{13}{2}\zeta(3)\right]
\nonumber\\
&\ & \hspace{0.1mm} 
+ \, \frac{1}{16}\, C_F T_F n_F\, \left[ \frac{65}{54} + \zeta(2)\right] \, ,
\nonumber\\
E^{[g] \, (2)}_1 
&=& 
\frac{1}{32} C_A^2 \, \left[ - \frac{346}{27} 
                           + \frac{11}{6} \zeta_2 + \zeta_3 \right]
+ \frac{1}{16} C_A T_F n_F \, \left[ \frac{64}{27} - \frac{\zeta_2}{3} \right] 
+ \frac{1}{16} C_F T_F n_F \, ,
\label{Enmvalues}
\eea
where for $E^{[i] \, (2)}_1$ we give the explicit expressions for the quark 
and gluon cases.
Notice that the full single pole term includes a contribution from the running
of the finite term at one loop, which appears as an ${\cal O}(\vep)$ contribution
in ${\cal G}^{[i] \, (1)}$.   

\section{Eikonal amplitudes at one and two loops}
\label{GammaScalc}

We begin this section with a calculation of the 
soft 
anomalous dimension matrix 
for quark-antiquark elastic scattering
at one and two 
loops, in terms of a specific color basis \cite{KOS},
and then discuss the representation of the matrix in the color generator
notation of \cite{catani96,catani98}.
We will see that the basic result of our calculation, the proportionality
of the one- and two-loop matrices, applies to a much wider
class of processes.

\subsection{$2 \rightarrow 2$ eikonal diagrams at one loop}

\begin{figure}[t]
\centerline{\epsfxsize=7cm \epsffile{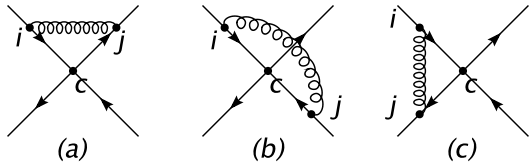}}
\caption{\label{Fi:one_loop}One-loop diagrams that contribute 
to $\Gamma_{S_\f}^{(1)}$}
\end{figure} 
Here we will present the calculation for one loop 
corrections  to $W$, Eq.~(\ref{eq:wivertex}), 
for quark-antiquark scattering, and by using 
Eq.~(\ref{Eq:gamma_oneloop}) we will derive the 
corresponding one-loop soft anomalous dimension matrix. 
Representative one-loop diagrams are shown in Fig.~\ref{Fi:one_loop}. 
One can write the amplitude for any diagram $D$ as
\begin{equation}
M_D=F_D\times C_{D,I} \,,
\end{equation}
where $F_D$ is the corresponding 
Feynman integral in dimensional regularization and 
$C_{D,I}$ is the color tensor.   
We will refer to $F_D$ as the velocity factor below, because it absorbs
all dependence on kinematic variables.  To uniquely define the normalizations
of the velocity factors, and hence the color tensors, we define them
to equal the corresponding integrals for the scattering of eikonal
lines that couple to the exchanged gluons
via color-independent ``abelian" vertices.   
In particular, we absorb into the velocity factors 
the $(-1)$ associated with a gluon coupled to an eikonal
line in the anti-quark representation.  
Note that this separation of color and velocity factors is possible
even if the eikonal lines are in the adjoint representation.
This method will facilitate our eventual comparison to 
results expressed in the formalism of Ref.~\cite{catani98}.

Consider the left-hand diagram in Fig.~\ref{Fi:one_loop}, which 
we will call a ``$t$-channel diagram", referring to the pair of
eikonal lines to which the gluon is connected.    
We will follow Refs.~\cite {KR87,KK95}, and express the velocity factor 
as an integral in configuration space.  For an arbitrary one-gluon
correction to a phase operator of the form of Eq.~(\ref{eq:wivertex}),
such a correction is given by
\begin{equation}
\label{Eq:one_loop_integral}
F_t=(ig\mu^{\vep})^2\int_{{\mathcal C}_i}
dx_{\mu}\int_{C_j}dy_{\nu}\,D^{\mu\nu}(x-y),
\end{equation}
where integration is performed over the positions of gluons 
on the paths of the Wilson lines, ${\mathcal C}_i$ and ${\mathcal C}_j$. 
For the lines in Eq.~(\ref{eq:wivertex}) these paths are specified by
\begin{eqnarray}
{\mathcal C}_i=v_i\,\beta, \qquad\quad
{\mathcal C}_j=v_j\,\alpha,
\end{eqnarray}
where $\alpha$ and $\beta$ run from $-\infty$ to $0$ ($0$ to $\infty$) for
an incoming (outgoing) path.  
For the $t$-channel diagram shown in Fig.~\ref{Fi:one_loop}$(a)$, 
for example, where $t = (p_1-p_3)^2 = (p_2-p_4)^2$, we may have 
$\{i,j\}=\{1,3\}$ or $\{2,4\}$.

In Feynman gauge the coordinate-space gluon propagator, 
in dimensional regularization with $D=4-2\vep$, 
is given by~\cite{KK95}
\begin{eqnarray}
D_{\mu\nu}(x)&=&g_{\mu\nu}D(x)\nonumber\\
&=&g_{\mu\nu}\frac{\Gamma(1-\vep)}{4\pi^{2-\vep}}
\,\frac{1}{(x^2-i\epsilon)^{1-\vep}}\,.
\label{Dxdef}
\end{eqnarray}
Using this expression in Eq.~(\ref{Eq:one_loop_integral}), we have
\begin{eqnarray}
F_t&=&(ig\mu^{\epsilon})^2 \int_0^{\infty}d\alpha\int_{-\infty}^0d\beta\,
v_i^\mu D_{\mu\nu}(v_j\alpha-v_i\beta)v_j^\nu \nonumber\\
&=&(ig\mu^{\vep})^2(v_i\cdot v_j)\frac{\Gamma(1-\vep)}{4\pi^{2-\vep}}
\int_0^{\infty}d\alpha\int_0^{\infty}d\beta\frac{1}{\left[(v_j\alpha+v_i\beta)^2-i\epsilon\right]^{1-\vep}}\, .
\label{Ftintegral}
\end{eqnarray}
As observed above, all such integrals vanish in dimensional regularization,
since they are scaleless.  The contribution of each such velocity-dependent 
integral is given by its counterterm, equal to its infrared pole
and hence to the negative of its ultraviolet (UV) pole. 
Of course, $F_t$ may be evaluated as a momentum-space integral with
equivalent results.  

In order to isolate the (single) UV pole in Eq.~(\ref{Ftintegral}),
we apply an infrared cut-off for the integral 
by introducing a small parameter $\lambda$ with units of mass.  
This can be effected simply by inserting $\theta(1/\lambda - \alpha)$
in Eq.~(\ref{Ftintegral}).  The $\alpha$ and $\beta$ integrals are then 
easily related to a single integral in terms of 
$z=\frac{\alpha}{(\alpha+\beta)}$ (see Eq.~(\ref{FtNewVariables}) in
Appendix~\ref{OneloopVelFactAppendix}). We find 
\begin{equation}
F_t=-\left(\frac{\alpha_s}{\pi}\right)\left(\frac{\pi\mu^2}{\lambda^2}\right)^{\vep}\Gamma(1-\vep)\frac{1}{2\vep}
\int_0^1dz\frac{v_i\cdot v_j}{\left([v_jz+v_i(1-z)]^2\right)^{1-\vep}}
\,.
\label{Ftzintegral}
\end{equation}
The single UV pole term in this expression is given by~\cite{KR87,KK95}
\begin{equation}
F_t^{s.p.}(v_i,v_j)=- \left(\frac{\alpha_s}{\pi}\right)\frac{1}{2\vep}\gamma_{ij}\coth\,\gamma_{ij}\, ,
\end{equation}
where 
\begin{equation}
\cosh\,\gamma_{ij}\equiv \frac{v_i\cdot v_j}{\sqrt{v_i^2v_j^2}}\, . 
\end{equation}
Because there is only a single, overall infrared divergence
in $F_t$, any such cut-off will give the same ultraviolet pole.

In the high-energy limit ($\gamma_{ij}\gg1$), we have
\begin{equation}
F_t^{s.p.}(v_i,v_j)=-\left(\frac{\alpha_s}{\pi}\right)\frac{1}{2\vep}\gamma_{ij}
\,.
\label{Ftspvivj}
\end{equation}
For $\{i,j\}=\{1,3\}$ and $\{i,j\}=\{2,4\}$ the answers are identical, 
in this $2 \rightarrow 2$ process. In the high-energy limit we define
\begin{equation}
\gamma_{13}=\gamma_{24}=T \,, \qquad\quad
\gamma_{14}=\gamma_{23}=U \,, \qquad\quad
\gamma_{12}=\gamma_{34}=S\, ,
\end{equation}
where 
\begin{eqnarray}
T &=& \ln\left(\frac{2v_1\cdot v_3}{v^2}\right)
=\ln\left(\frac{2v_2\cdot v_4}{v^2}\right)\, ,\nonumber\\
U &=& \ln\left(\frac{2v_1\cdot v_4}{v^2}\right)
=\ln\left(\frac{2v_2\cdot v_3}{v^2}\right) \, ,\nonumber\\
S &=& \ln\left(\frac{-2v_1\cdot v_2}{v^2}\right)
=\ln\left(\frac{-2v_3\cdot v_4}{v^2}\right) \, ,
\label{TUSdefs}
\end{eqnarray}
with $v_i^2\equiv v^2$ for all $i$. The velocity 
factors for $u$- and $s$-channel diagrams are found by
taking into account the extra minus sign associated with coupling to an
eikonal line in the antiquark representation, 
as well as 
that 
from crossing substitutions, which change the sign of 
$\coth \gamma_{ij}$ from unity to $-1$ in the high-energy limit,
\begin{eqnarray}
F_u(v_i,v_j) &=&  -\; F_t(v_i,v_j) \, , \nonumber\\ 
F_s(v_i,v_j) &=&   F_t(v_i,-v_j)\, . \label{Fsdef}
\end{eqnarray}
Here, $\{i,j\}=\{1,4\}$ and $\{2,3\}$ for the $u$-channel diagrams, and 
$\{i,j\}=\{1,2\}$ and $\{3,4\}$ for the $s$-channel diagrams.   
The function $F_s$ has the same overall sign as $F_t$
because it differs both by an antiquark connection and by crossing, while
$F_u$ has the opposite sign.
\begin{figure}[t]
\centerline{\epsfxsize=7cm \epsffile{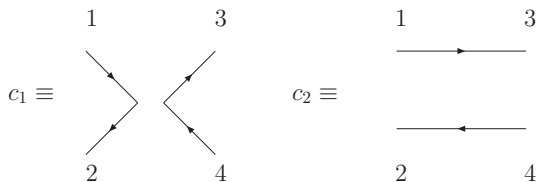}}
\caption{\label{Fi:color_basis}Color basis $\{c_1,c_2\}$ for four-quark process}
\end{figure} 

In summary, the single poles for
the velocity factors for the diagrams in Fig.~\ref{Fi:one_loop} are given by
\begin{eqnarray}
F_t^{s.p.} &=& - \left(\frac{\alpha_s}{\pi}\right)\frac{1}{2\vep}T \, , \nonumber\\
F_u^{s.p.} &=&  \left(\frac{\alpha_s}{\pi}\right)\frac{1}{2\vep}U \, , \nonumber\\
F_s^{s.p.} &=& - \left(\frac{\alpha_s}{\pi}\right)\frac{1}{2\vep}S\, .
\label{1loopFs}
\end{eqnarray}
To construct the counterterms, of course, we must also compute 
the corresponding color tensors for each diagram.

We will use $C^{[a]}_i$ to denote the color tensor for the $t$-channel diagram 
in Fig.~\ref{Fi:one_loop}, with color tensor $c_i$, $i=1,2$
at short distances.   For the latter we choose the basis tensors
shown in Fig.~\ref{Fi:color_basis}.  
The coefficients of the color tensors absorb all overall factors
not included in the velocity factors of Eq.~(\ref{1loopFs}).

One can calculate these color tensors from the basic identity for 
the generators of SU$(N_c)$,
\begin{equation}
\sum_a\,(T^a)_{r_2r_1}\,(T^a)_{r_3r_4}
=\frac{1}{2}\delta_{r_2r_4}\delta_{r_3r_1}
-\frac{1}{2N_c}\delta_{r_2r_1}\delta_{r_3r_4}\, .
\label{gluesplit}
\end{equation}
In the color basis given in Fig.~\ref{Fi:color_basis}, the color tensors of the 
$t$-channel diagrams are given by
\begin{eqnarray}
C_1^{[t]}&=&-\frac{1}{2N_c}c_1+\frac{1}{2}c_2 \, ,
\nonumber\\
C_2^{[t]}&=& \frac{N_c^2-1}{2N_c} \, c_2 = C_F\,c_2\, .
\label{CtsubI}
\end{eqnarray}
We will employ a similar notation below for other one-loop and 
for two-loop diagrams. Color tensors for the $u$ and $s$ channel 
diagrams are computed in a similar way with the results
\begin{eqnarray}
C_{1}^{[u]}&=&  - \frac{1}{2N_c}c_1 + \frac{1}{2}c_2 \,,\nonumber\\
C_{2}^{[u]}&=&  \frac{1}{2}c_1 - \frac{1}{2N_c}c_2 \,,
\label{CusubI}
\end{eqnarray}
and
\begin{eqnarray}
C_{1}^{[s]}&=&  C_F\,c_1  \,, \nonumber\\
C_{2}^{[s]}&=& \frac{1}{2}c_1 - \frac{1}{2N_c}c_2 \,.
\label{CssubI}
\end{eqnarray}
We summarize these relations in matrix form by
\bea
C_I^{[a]} = \sum_{J=1,2} c_J\, d^{[a]}_{JI} \,,
\label{colornotation}
\eea
where the matrix element $d^{[a]}_{JI}$ specifies the mixing from
color tensor $c_I$ to tensor $c_J$ by the exchange of a gluon
in channel $a=t,u,s$.  An important identity that we will use below is 
\bea
 d^{[t]}_{JI} + d^{[s]}_{JI} - d^{[u]}_{JI} = C_F\, \delta_{JI}\, ,
\label{ddiagonal}
\eea
which we easily verify from the relations above.
Note that this equality holds in an arbitrary representation.
\footnote{Equation~(\ref{ddiagonal}) is equivalent
in this case to the well-known
identity $\sum_i {\bf T}_i=0$ in the color generator 
notation that we will review below.}

The contribution of each diagram to the matrix counterterm 
$Z_{W_\f}^{(1)}$ is now found by
the product of the corresponding color factor times
the pole part of the velocity factors,
\bea
\left(Z_{W_\f}^{(1)}\right)_{JI} = 2\, \sum_{a=s,t,u} d^{[a]}_{JI}
\, \frac{F_a^{s.p.}}{\alpha_s/\pi} \, ,
\label{introdJI}
\eea
in terms of the single-pole terms of Eq.~(\ref{1loopFs}) and the color factors
read off from Eqs.~(\ref{CtsubI})--(\ref{CssubI}).
Given the counterterm matrix, we can evaluate 
${\bf\Gamma}_{W_\f}^{(1)}$ by using Eq.~(\ref{Eq:gamma_oneloop}) 
with the result 
\begin{equation}
\label{W1loop}
{\bf\Gamma}_{W_\f}^{(1)}
=\left( \begin{array}{cc} \frac{1}{N_c}\left(U-T\right)+2\,C_F\,S &
\left(S-U\right)\\ \\ \left(T-U\right) &\frac{1}{N_c}\left(U-S\right)+2\,C_F\,T
\end{array} \right).
\end{equation}\\
Exactly the same calculation gives $\Gamma_2^{[i]}$, 
the anomalous dimension for the eikonal jet function, 
defined as the square root of the eikonal singlet form factor,
Eq.~(\ref{Wiidef}).  In Eq.~(\ref{Ftspvivj}), we simply
let  $\cosh\,\gamma_{ij} \to u_0/v^2 = \mu^2/(Q^2v^2)$ (using Eq.~(\ref{u0def})), 
in the limit $v^2 \to 0$.  The one-loop result for parton $i$ is then
given by
\begin{equation}
\Gamma_2^{[i]\,(1)}
\left(\frac{u_0}{v^2}\right)
= \frac{1}{2}\, C_i\, \ln\left(\frac{\mu^2}{Q^2v^2}\right),
\end{equation}
with $C_i=C_F$ for quarks and $C_A$ for gluons. By using this 
expression and the definition for $\Gamma_{S_\f}$, Eq.~(\ref{GammaSIJdef}),
we find \\
\begin{equation}
\label{Eq:1l_gammaS}
{\bf \Gamma}_{S_\f}^{(1)}=\left( \begin{array}{cc} \frac{1}{N_c}\left(\mathcal{U}-\mathcal{T}\right)+2\,C_F\,\mathcal{S} &
\left(\mathcal{S}-\mathcal{U}\right)\\ \\ \left(\mathcal{T}-\mathcal{U}\right) &\frac{1}{N_c}\left(\mathcal{U}-\mathcal{S}\right)+2\,C_F\,\mathcal{T}
\end{array} \right),
\end{equation}\\ 
where
\begin{equation}
\mathcal{T}\equiv\ln\left(\frac{-t}{\mu^2}\right) \,, \qquad\quad
\mathcal{U}\equiv\ln\left(\frac{-u}{\mu^2}\right) \,, \qquad\quad
\mathcal{S}\equiv\ln\left(\frac{-s}{\mu^2}\right)\,.
\label{calTUSdefs}
\end{equation}
After performing the subtraction of the jet functions, we set
$v_i \to \beta_i$, and then use Eqs.~(\ref{betaidef}) and (\ref{u0def})
to recast the result in terms of the usual Mandelstam variables,
$t$, $u$ and $s$.
We notice that, as anticipated, all collinear logarithms, and
hence sensitivity to our choice of collinear regulation,
are absent in the soft anomalous dimension matrix, ${\bf \Gamma}_{S_\f}^{(1)}$.

\subsection{$2 \rightarrow 2$ eikonal diagrams at two loops}

\begin{figure}[ht]
\centerline{\epsfxsize=8cm \epsffile{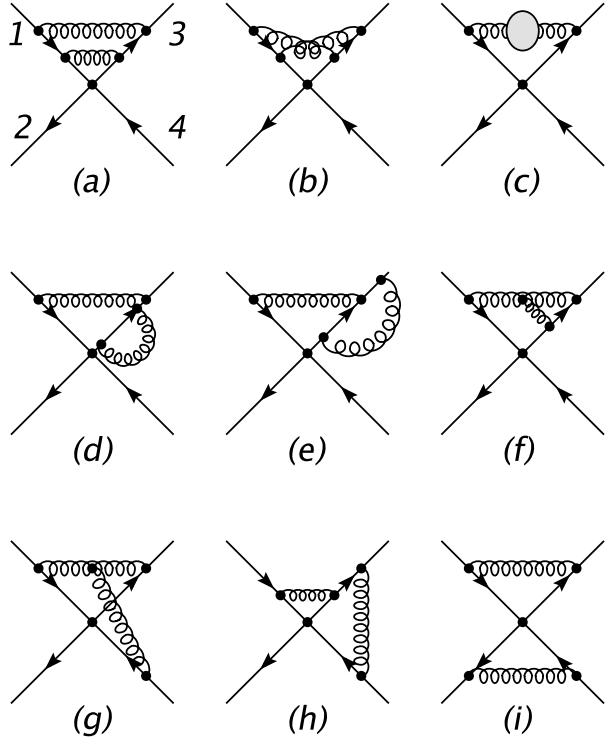}}
\caption{\label{Fi:2ld}Two-loop diagrams that contribute to 
$\Gamma_{W_\f}^{(2)}$}
\end{figure}
Figure~\ref{Fi:2ld} shows the classes of topologically 
inequivalent diagrams that contribute to $\Gamma_{S_\f}^{(2)}$, 
when combined with their one-loop counterterm diagrams. 
One obtains the full set from all different combinations 
of external legs with these topologies. It is easy to see 
that the number of graphs for each inequivalent set is 
$N_a=6$, $N_b=6$, $N_c=6$, $N_d=12$, $N_e=12$, $N_f=12$, $N_g=4$, $N_h=24$ and 
$N_i=3$, which in total gives $85$ two-loop diagrams.  
As in the one-loop case, we find anomalous dimensions from
the combinations of velocity factors and color tensors.   

Consider first diagram $(i)$, which is the only two-loop topology
involving all four eikonal lines.
Diagram $(i)$ does not have a 
surviving single UV pole when we add its one-loop counterterms. 

Regarding the remaining cases, we consider first those diagrams 
involving two eikonal lines only, which we refer to as ``2E" diagrams.
Next, we will show that diagram $(g)$ vanishes, which we 
consider a very important result.  Finally, we will calculate 
all the contributions from the surviving ``3E'' diagram type $(h)$.
In this case, we will find that the diagrams, although 
non-vanishing, reduce to the product of one-loop diagrams,
and thus do not contribute to the two-loop anomalous dimensions.

\subsubsection{The 2E diagrams and $\Gamma_i^{(2)}$}

In the 2E diagrams, $(a)$, $(b)$, $(c)$, $(d)$, $(e)$ and $(f)$, 
the gluons connect to only two of the four eikonal lines in $W$.  These
same diagrams also contribute to the two-loop cusp anomalous dimension,
$\Gamma_2^{[i]\,(2)}$, and their single UV poles are well known~\cite{KK95}. 
We review their velocity factors here, because they
are needed for the two-loop anomalous dimension matrix.
Additional details are given in Appendix~\ref{VelFact2ESubAppendix}.

The color factors of the 2E diagrams 
are proportional to the color-mixing matrix elements 
for single $t$-channel gluon exchange, $d^{[t]}_{JI}$, 
defined in Eq.~(\ref{colornotation}).
This is manifestly the case for the individual 
diagrams $(c)$, $(d)$, $(e)$ and $(f)$.
For the sum of diagrams $(a)$ and $(b)$, it relies on the 
result~\cite{KR87,KK95} that the single-pole terms in 
the velocity factors of these two diagrams are the negatives
of each other.  The net color factor for the $(a)$ and $(b)$
single poles is then proportional to
the commutator of two generators, 
which allows it to be expressed in terms of the one-loop color
factor, as $C_A \, d^{[t]}_{JI}$.
We can thus present the contributions of all the 2E
diagrams in terms of the $d^{[a]}_{JI}$, with $a=s,t,u$.

In terms of the factors $d^{[t]}_{JI}$, the 
two-loop counterterms\footnote{These results, of
 course, require that we combine these diagrams with
 the corresponding one-loop counterterms for their divergent subdiagrams.
 Notice that diagram $(b)$ in Fig.~\ref{Fi:2ld} does not have a one-loop 
 counterterm since it does not have a divergent subdiagram.} 
 for the diagrams $(a)$, $(b)$, $(c)$ and $(f)$ in the high-energy limit are,
analogously to the one-loop velocity factors, Eq.~(\ref{1loopFs}),
\bea
\left( Z_{W_\f}^{(a+b)}\right)_{JI}
 &=&  
-  \left(\frac{\alpha_s}{\pi}  \right)^2\, d^{[t]}_{JI}
\, \frac{C_A}{2}\, \frac{1}{2\vep}\,
 \left [ \frac{T^3}{6} +  \frac{\zeta(2)}{2} \, T
        - \frac{\zeta(3)}{2} \right ]\, , 
\nonumber\\ 
\, \nonumber\\
\left( Z_{W_\f}^{(c)}\right)_{JI}
 &=& - \left(\frac{\alpha_s}{\pi}  \right)^2\, d^{[t]}_{JI}
\, \frac{1}{2}\, \frac{1}{2\vep}\,
 \left(\, \frac{31}{36} C_A - \frac{5}{9} T_F n_F \, \right)\,T\, ,
\label{Fabc2loops}
\eea
and
\begin{eqnarray}
\left( Z_{W_\f}^{(f)}\right)_{JI}
&=&  - \left(\frac{\alpha_s}{\pi}  \right)^2
\, d^{[t]}_{JI}\, \frac{C_A}{2}\,
 \frac{1}{4\vep}\Bigg\{\, \left[ - \frac{T^3}{6} 
   + \left(1 -  \zeta(2) \right) T  \right]
\nonumber\\
&\ & \hspace{40mm} + \left[\frac{T^2}{2}-T+\zeta(2) \right]\, \Bigg \}\, ,
\label{Ff2loops}
\end{eqnarray}
with $T$ the logarithm of $2v_1\cdot v_3/v^2$, as in Eq.~(\ref{TUSdefs}).
In Eq.~(\ref{Ff2loops}), the second term in square brackets gives
the result of those numerator terms that are proportional to $v_3^2$
before the integration.
(See Appendix~\ref{VelFact2ESubAppendix}.)
The entire $T$-dependence of these terms cancels against
the contributions from diagrams $(d)$ and $(e)$, 
which are also proportional to $v_3^2$ before integration
and are given individually by
\bea
\label{Fde2loops}
\left( Z_{W_\f}^{(d)}\right)_{JI}
 &=&- \left(\frac{\alpha_s}{\pi}  \right)^2
\, d^{[t]}_{JI}\,C_F\,\frac{1}{4\vep}
\,\left[-\frac{T^2}{2}+T-\frac{\zeta(2)}{2}\right]\,,
 \nonumber\\
 \left( Z_{W_\f}^{(e)}\right)_{JI}
 &=&\left(\frac{\alpha_s}{\pi}  \right)^2
\, d^{[t]}_{JI}\,\left(C_F-\frac{C_A}{2}\right)
\,\frac{1}{4\vep}\,\left[-\frac{T^2}{2}+T-\frac{\zeta(2)}{2}\right]\,.
\eea
The combined $t$-channel contribution from the six diagrams 
$(a)$, $(b)$, $(c)$, $(d)$, $(e)$ and $(f)$ to the soft anomalous
dimension matrix is found by adding 
Eqs.~(\ref{Fabc2loops})--(\ref{Fde2loops})\footnote{%
Note that one needs to multiply Eqs.~(\ref{Ff2loops}) and 
(\ref{Fde2loops}) by 2 because for these diagrams there are two 
ways of attaching the gluons to the eikonal lines.},
\bea
\label{Eq:Gamma_cusp}
\left( Z_{W_\f}^{[t]}\right)_{JI}
&\equiv& 2 \left( Z_{W_\f}^{(a+b)} + Z_{W_\f}^{(c)}
+ 2 \Bigl[ Z_{W_\f}^{(d)} + Z_{W_\f}^{(e)} + Z_{W_\f}^{(f)} \Bigr]
\right)_{JI}
\nonumber\\ 
&=&  
- \frac{1}{2\vep}\,
\left(\frac{\alpha_s}{\pi}\right)^2\,  d^{[t]}_{JI}\,
\Bigg\{ \left[ C_A \,\left(\frac{67}{36}-\frac{\zeta(2)}{2}\right) -
\frac{5}{9} T_F n_F \right]\, 
\ln\left(\frac{2v_1\cdot v_3}{v^2}\right)
\nonumber\\
&& \hskip33mm
+ \, \frac{C_A}{2} \, \left(\zeta(2)-\zeta(3)\right) \Bigg\}
\nonumber\\
&=& \!    \left(\frac{\alpha_s}{\pi}\right)^2\,
d^{[t]}_{JI}\, \left \{ \,  \frac{K}{2}
\,   \frac{F_t^{s.p.}}{(\as/\pi)}\,  
- \frac{C_A}{4\vep}\, \left(\zeta(2) - \zeta(3)\right) \right \}  \, ,
\eea
where the second line recalls a standard 
notation~\cite{KodairaTrentadue}
for the quantity $K$,  
\begin{equation}
K \equiv C_A \,\left(\frac{67}{18} - \zeta(2)\right)
 - \frac{10}{9} T_F n_F \,.
\label{KDef}
\end{equation}
The result~(\ref{Eq:Gamma_cusp}) includes a factor of two for
the other $t$-channel exchange, between lines 2 and 4.

Analogous considerations, of course,
apply to diagrams with pairs of $s$- and $u$-channel 2E diagrams.
Together with the $t$-channel diagrams, they contribute
to the two-loop anomalous dimension matrix for $W$
according to Eq.~(\ref{Eq:gamma_twoloops}),
\bea
{\bf \Gamma}^{(2E)}_{W_\f}{}^{(2)}
&=&
\frac{K}{2} \sum_{i=s,t,u} d^{[i]}_{JI}\, 
\left(\, \frac{-2\vep F_i^{s.p.}}{(\as/\pi)}\, \right)
+ \delta_{JI}\, C_A \, C_i \left(\zeta(2)-\zeta(3)\right)
\nonumber\\
&=& \frac{K}{2}\, {\bf \Gamma}_{W_\f}^{(1)} + 
\delta_{JI}\, C_A \, C_i \left(\zeta(2) -\zeta(3)\right)\, ,
\label{Eq:GammaWfinal}
\eea
where we have used the identity~(\ref{ddiagonal}),
and where ${\bf \Gamma}_{W_\f}^{(1)}$ is the
same one-loop anomalous dimension given
in Eq.~(\ref{W1loop}).

In a precisely similar manner we find for the two-loop
form factor (cusp) anomalous dimension for partonic 
representation $i$,
\bea
\Gamma_2^{[i]\,(2)}
\left(\frac{u_0}{v^2}\right)
= \frac{C_i}{4}\, \left[
 K\, \ln\left(\frac{\mu^2}{Q^2v^2}\right) 
+ C_A \, \left(\zeta(2)-\zeta(3)\right) \right ]\, .
 \label{Eq:Gammai2}
\eea
As at one loop, we combine Eqs.~(\ref{Eq:GammaWfinal}) 
and (\ref{Eq:Gammai2}) in Eq.~(\ref{GammaSIJdef}),
in order to find the contribution of the 2E diagrams to the
two-loop soft anomalous dimensions for scattering.
 
It is now clear that the two-loop soft anomalous
dimension matrix inherits from the 2E diagrams a factor of 
$K$ times the one-loop anomalous dimension matrix.  
The result is,   
\bea
{\bf \Gamma}^{(2E)}_{S_\f}{}^{(2)} 
= \frac{K}{2}\, {\bf \Gamma}_{S_\f}^{(1)}\, ,
\label{Gamma2E}
\eea
with ${\bf \Gamma}_{S_\f}^{(1)}$ the same one-loop anomalous 
dimension given in Eq.~(\ref{Eq:1l_gammaS}).  All velocity-independent
terms in the 
$Z_{W_\f}^{(2E)}$ 
that are not in ${\bf \Gamma}_{S_\f}^{(1)}$
cancel 
in Eq.~(\ref{GammaSIJdef}) against the
corresponding finite terms from the eikonal form factors
in the two-loop soft
anomalous dimension, along with all collinear-singular
dependence.   This is important, because the constant terms
depend in general on the eikonal approximation and our choice of 
collinear regularization.  At the same time, we have now used 
{\it all} the collinear-singular dependence in
the Sudakov anomalous dimensions, Eq.~(\ref{Eq:Gammai2}), to cancel
the $\ln v^2$-dependence of the 2E diagrams of $W$.  The 3E diagrams,
represented by $(g)$ and $(h)$ in Fig.~\ref{Fi:2ld}, 
have no remaining subtractions.  The combination of
these classes of diagrams must therefore be 
free of collinear singularities at the two-loop level.  

\subsubsection{Vanishing of three-gluon diagram with three eikonal lines}

Now let's show that diagram $(g)$ in Fig.~\ref{Fi:2ld} vanishes. 
Up to overall factors 
which play no role, the velocity Feynman 
integral for a generic three-gluon diagram can be written as
\begin{eqnarray}
F(v_A,v_B,v_C)
&=&
\int d^Dk_1 d^Dk_2\, \frac{1}{v_B\cdot k_1+i\epsilon}\,\frac{1}{v_A\cdot k_2+i\epsilon}\,\frac{1}{v_C \cdot \left(k_1+k_2\right)+i\epsilon}
\,\frac{1}{k_1^2+i\epsilon}\,\frac{1}{k_2^2+i\epsilon}\nonumber\\ 
&\ & \hspace{-33mm}\times \frac{1}{(k_1+k_2)^2+i\epsilon}\times
\Bigg[ v_A\cdot v_B \, v_C\cdot \left(k_1-k_2\right)
+
v_A\cdot v_C\, v_B\cdot \left(  k_1 + 2k_2 \right)
+
v_B\cdot v_C\, v_A \cdot \left( -2k_1 - k_2\right)\, \Bigg],
\nonumber\\
\end{eqnarray}
where the term in square brackets is the three-gluon
vertex momentum factor. Here $v_A$, $v_B$ and $v_C$ are three different 
eikonal velocities. We take lightlike $v_A^2=v_B^2=0$.
We can then expand any momentum $p^\mu$ as
\begin{equation}
p^\mu = \frac{v_A^\mu}{v_A\cdot v_B} \, v_B\cdot p
\, + \,
\frac{v_B^\mu}{v_A\cdot v_B} \, v_A\cdot p
\, + \, p_{T}^\mu\, ,
\end{equation}
with $p_T^\mu$ the transverse components, satisfying 
$v_A\cdot p_T = v_B\cdot p_T = 0$.

For use in the integral we introduce the variables:
\begin{eqnarray}
\xi_i =  \frac{v_A\cdot v_C}{v_A\cdot v_B} \, v_B\cdot k_i\, ,
\qquad \qquad
\eta_i = \frac{v_B\cdot v_C}{v_A\cdot v_B} \, v_A\cdot k_i\, .
\label{xieta}
\end{eqnarray}
We introduce these variables into the integral
by using $v_A$ and $v_B$ to define light-cone coordinates,
\begin{eqnarray}
dk_i^+dk_i^- &=& \frac{1}{v_A\cdot v_B}\, d(v_B\cdot k_i)\, d(v_A\cdot k_i)
\nonumber\\
&=&
\frac{v_A\cdot v_B}{(v_A\cdot v_C)(v_B\cdot v_C)}\, d\xi_i d\eta_i\, ,
\end{eqnarray}
so that
  \begin{eqnarray}
k^2_i = 2\frac{(v_A\cdot k_i)(v_B\cdot k_i)}{v_A\cdot v_B} - k_{i,T}^2
\ = \ 
  2 \frac{v_A\cdot v_B}{(v_A\cdot v_C)(v_B\cdot v_C)}\, \xi_i\eta_i - k_{i,T}^2\, .
\end{eqnarray}

When we change variables in $F$ to the $\xi$'s and $\eta$'s, we find
\begin{eqnarray}
F(v_A,v_B,v_C)
&=&
\frac{ v_A\cdot v_B }{ (v_A\cdot v_C) (v_B\cdot v_C) } \, 
\int\, \prod_{i=1}^2\,  d\xi_i\, d\eta_i\, d^{D-2}k_{i,T}\
\frac{1}{2\frac{v_A\cdot v_B}{(v_A\cdot v_C)(v_B\cdot v_C)}\, \xi_i\eta_i - k_{i,T}^2+i\epsilon}
\nonumber\\
&\ & \hspace{5mm} \times 
\frac{1}{2\frac{v_A\cdot v_B}{(v_A\cdot v_C)(v_B\cdot v_C)}\, (\xi_1+\xi_2)(\eta_1+\eta_2)
- \left( k_{1,T}+k_{2,T}\right)^2+i\epsilon}
\nonumber\\
&\ &  \hspace{5mm} \times  
\ \frac{1}{\xi_1+i\epsilon}\frac{1}{\eta_2+i\epsilon}\,  
\frac{1}{ \xi_1+\xi_2 +  \eta_1+\eta_2 
- v_{C,T}\cdot \left(k_1+k_2\right)_T+i\epsilon}
\nonumber\\
  &\ & \hspace{5mm}\times 
\bigg[ \xi_1 - \xi_2 + \eta_1 - \eta_2
- v_{C,T} \cdot \left(k_1-k_2\right)_T 
+ \xi_1 + 2\xi_2 - 2\eta_1 -  \eta_2 \, \bigg] 
\nonumber\\
&=& 0 \,.
\end{eqnarray}
The integral vanishes because the numerator is antisymmetric 
under $\xi_1 \leftrightarrow \eta_2$,
$\xi_2 \leftrightarrow \eta_1$ and $k_{1,T} \leftrightarrow k_{2,T}$, 
while the product of the denominators is symmetric. 
Notice that group factors play no part in this argument. 
This result is therefore very general and applies to any 
$2\rightarrow n$ process with lightlike velocities.

\begin{figure}[t]
\centerline{\epsfxsize=7cm \epsffile{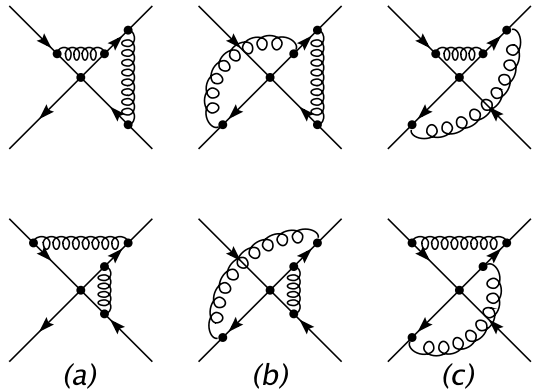}}
\caption{\label{Fi:3E_pairs} 
a-c) Pairs of  3E diagrams.}
\end{figure}

\subsubsection{Exponentiation of the remaining 3E diagrams}

The only remaining class of diagrams is illustrated by diagram $(h)$ 
in Fig.~\ref{Fi:2ld}. 
Along with its companions found by permuting the eikonal lines, 
we can refer to these as ``3E'' diagrams, since 
they
are the only nonvanishing diagrams with gluons connected to three 
eikonal lines. 
There are $24$ such 3E diagrams, $8$ of them with 
$s$ and $t$ channel gluon exchanges, 
$8$ of them with $s$ and $u$ channel gluons, 
and finally $8$ of them with $t$ and $u$ channel gluons.  
They come in pairs as shown in Fig.~\ref{Fi:3E_pairs}. 

We now show that the analysis of the previous subsection
regarding the three-gluon diagrams leads to a very interesting
result for the remaining 3E diagrams as well.  In this case, the diagrams
do not vanish, but reduce to products of one-loop 
diagrams\footnote{The reduction of a different class of 
multi-loop eikonal diagrams, namely the 2E diagrams of ladder type, 
to powers of one-loop diagrams, was previously observed in 
Ref.~\cite{KidonakisExp} to hold to all loop orders.}.
They therefore provide no contribution to the two-loop anomalous
dimension matrix.

Each such 3E diagram contains one eikonal line with two gluons attached 
to it, which we label as $v_C$.  The two lines having one gluon attached
are labelled $v_A$ and $v_B$.  We will consider a pair of 3E diagrams 
that are related simply
by exchanging the order in which the two gluons attach to $v_C$,
as in Fig.~\ref{Fi:3E_pairs}$(a)$ for example.
The two diagrams have differing color and momentum structures, but we can rewrite
their sum as the sum of one term with symmetric color and momentum integrals,
plus a second term with antisymmetric color and momentum integrals.  
In the spirit of the discussion above, we suppress the color matrices 
held in common and write
\bea
{\cal F}_{ab}(v_A,v_B,v_C) \equiv {\cal F}_{ab}^{({\rm sym})}(v_A,v_B,v_C)
 + {\cal F}_{ab}^{({\rm antisym})}(v_A,v_B,v_C)\, ,
 \eea
where the subscripts $a$ and $b$ refer to the color generators 
on the $v_C$-eikonal, contracted with generators on the 
$v_A$ or $v_B$ line, respectively.
Consider first the antisymmetric term, which is given by
\bea
{\cal F}_{ab}^{({\rm antisym})}(v_A,v_B,v_C)
&=&
\frac{1}{2}\, \left(T_b\, T_a - T_a\, T_b\right)\; 
\nonumber\\
&\ & \times
\int d^Dk_1 d^Dk_2\, \frac{1}{v_B\cdot k_1+i\epsilon}\,\frac{1}{v_A\cdot k_2+i\epsilon}\, 
\frac{1}{k_1^2+i\epsilon}\,\frac{1}{k_2^2+i\epsilon}\nonumber\\ 
&\ & \times
\frac{1}{v_C \cdot \left(k_1+k_2\right)+i\epsilon}\, 
\left[\,  \frac{1}{v_C \cdot k_1+i\epsilon}\, - \frac{1}{v_C \cdot k_2+i\epsilon} \right]\, .
\eea
(In the color-generator notation described in section~\ref{iterative} below,
the color operator associated with this antisymmetric term takes the form
$[ {\bf T}_B \cdot {\bf T}_C \, , \, {\bf T}_C \cdot {\bf T}_A]$.)
The same change of variables, Eq.~(\ref{xieta}), leads to an
expression that is again manifestly antisymmetric 
under the relabelling $\xi_{1,2}\leftrightarrow \eta_{2,1}$,
$k_{1,T}\leftrightarrow k_{2,T}$,
\bea
{\cal F}_{ab}^{({\rm antisym})}(v_A,v_B,v_C)
&=&
\frac{1}{2}\, \left(T_b\, T_a - T_a\, T_b\right)\; 
\nonumber\\
&& \hspace{-2mm} \times
\frac{ 1 }{ (v_A\cdot v_C) (v_B\cdot v_C) } \, 
\int\, \prod_{i=1}^2\,  d\xi_i\, d\eta_i\, d^{D-2}k_{i,T}\
\frac{1}{2\frac{v_A\cdot v_B}{(v_A\cdot v_C)(v_B\cdot v_C)}\, \xi_i\eta_i - k_{i,T}^2+i\epsilon}
\nonumber\\
&\ &  \hspace{-2mm} \times  
\ \frac{1}{\xi_1+i\epsilon}\frac{1}{\eta_2+i\epsilon}\,  
\frac{1}{ \xi_1  + \eta_1 
- v_{C,T}\cdot k_{1,T}+i\epsilon}\,
\frac{1}{ \xi_2 +  \eta_2 
- v_{C,T}\cdot k_{2,T}+i\epsilon}
\nonumber\\
  &\ & \hspace{-2mm}\times 
\frac{\xi_2 - \xi_1 + \eta_2 - \eta_1
- v_{C,T} \cdot \left(k_2-k_1\right)_T }
{ \xi_1+\xi_2 +  \eta_1+\eta_2 
- v_{C,T}\cdot \left(k_1+k_2\right)_T+i\epsilon}
\nonumber\\
&=& 0 \,.
\end{eqnarray}
The entire color-antisymmetric part of the infrared region thus
vanishes whenever the eikonal approximation is valid,
and the cancellation is exact for the eikonal
amplitudes we consider here.

Turning to the symmetric term, we need
only use the eikonal identity $1/[x(x+y)] + 1/[y(x+y)] = 1/(xy)$
to rewrite it as the product of the two lowest-order 
single-gluon exchange diagrams,
\bea
{\cal F}_{ab}^{({\rm sym})}(v_A,v_B,v_C)
&=&
\frac{1}{2}\, \left(T_b\, T_a + T_a\, T_b\right)\; 
\nonumber\\
&\ & \times
\int d^Dk_1 \frac{1}{v_B\cdot k_1+i\epsilon}\,
  \frac{1}{v_C \cdot k_1+i\epsilon}\, 
\frac{1}{k_1^2+i\epsilon}\,\nonumber\\ 
&\ & \times \int d^Dk_2\, \frac{1}{v_A\cdot k_2+i\epsilon}\, 
\frac{1}{v_C \cdot k_2+i\epsilon} \, \frac{1}{k_2^2+i\epsilon}\,  .
\eea
(In the color-generator notation described in section~\ref{iterative},
the color operator associated with this symmetric term takes the form
$\{ {\bf T}_B \cdot {\bf T}_C \, , \, {\bf T}_C \cdot {\bf T}_A \}$.)
These diagrams have the correct color and kinematic structure to 
represent the two-loop terms in the exponentiation
of the color matrix of one-loop infrared poles. 
They are precisely cancelled in the two-loop soft function 
by the corresponding products of one-loop counterterms.

We conclude that {\it the entire two-loop anomalous dimension is 
due to the 2E diagrams and is given
by Eq.~(\ref{Gamma2E})}, in which we may remove the
superscript $(2E)$, to obtain
\bea
{\bf \Gamma}_{S_\f}^{(2)} 
= \frac{K}{2}\, {\bf \Gamma}_{S_\f}^{(1)}\, .
\label{Gamma2complete}
\eea
We have thus determined that the two-loop anomalous dimension
color-mixing matrix is related to the one-loop matrix by the same factor
that relates the one- and two-loop Sudakov anomalous dimensions,
$A(\as)$.  Evidently, the next-to-next-to-leading poles in
amplitudes with color exchange are generated by the
same exponentiation of  ``webs" as for the elastic form factor
\cite{webs,Berger:2002sv}.
Additionally, we note that in the ``bremsstrahlung" or CMW
scheme~\cite{CMW}, this contribution, along with the corresponding
term in the cusp anomalous dimension, is absorbed into 
a redefinition of the strong coupling, 
which effectively boosts the strength of parton showering.

\subsection{Expansion of the soft function}

To relate the soft anomalous dimension to fixed-order 
calculations, we expand the resummed soft function, 
given as a path-ordered exponential in Eq.~(\ref{expoS}),
to order $\mathcal{O}(\alpha_s^2)$. The result is
\begin{eqnarray}
{\bf S}_\f\left(\frac{Q^2}{\mu^2}
=1,\alpha_s(Q^2),\ep \right)&=&
1+\frac{1}{2\vep}\left(\frac{\alpha_s}{\pi}\right){\bf \Gamma}_{S_\f}^{(1)}\,
+\frac{1}{8\vep^2}\left(\frac{\alpha_s}{\pi}\right)^2
\left({\bf \Gamma}_{S_\f}^{(1)}\right)^2\nonumber\\
&\ & - \frac{\beta_0}{16\vep^2}
\, \left(\frac{\alpha_s}{\pi}\right)^2{\bf \Gamma}_{S_\f}^{(1)}\,
+\frac{1}{4\vep}\,
\left(\frac{\alpha_s}{\pi}\right)^2
{\bf \Gamma}_{S_\f}^{(2)}
\nonumber\\
&=&
1+\frac{1}{2\vep}\left(\frac{\alpha_s}{\pi}\right){\bf \Gamma}_{S_\f}^{(1)}\,
+\frac{1}{8\vep^2}\left(\frac{\alpha_s}{\pi}\right)^2
\left({\bf \Gamma}_{S_\f}^{(1)}\right)^2\nonumber\\
&\ & - \frac{\beta_0}{16\vep^2}
\, \left(\frac{\alpha_s}{\pi}\right)^2{\bf \Gamma}_{S_\f}^{(1)}\,
+\frac{1}{8\vep}\,
\left(\frac{\alpha_s}{\pi}\right)^2\, K\, 
{\bf \Gamma}_{S_\f}^{(1)}
\, ,
\label{Sto2}
\end{eqnarray}
where in the second equality we have used Eq.~(\ref{Gamma2complete})
for the two-loop anomalous dimension matrix.
Combining this result with the second-order minimal jet function, 
Eq.~(\ref{logJmin}), in the formula for the factorized amplitude,
Eq.~(\ref{facampmin}), we will derive a result to compare directly
with the pole structure of explicit two-loop calculations.  

\subsection{Iterative color-matrix form}
\label{iterative}

Given the one- and
two-loop anomalous dimension soft matrices~(\ref{Eq:1l_gammaS}) and
(\ref{Gamma2complete}), and the expansion of the quark
jet function, as in Eq.~(\ref{logJmin}), we can use the factorized
amplitude, Eq.~(\ref{facampmin}), to calculate all
infrared and collinear poles at order $\as^2$ for quark-antiquark
scattering.  The result will be a set of coefficients of the specific
basis tensors in color space that we have chosen, Fig.~\ref{Fi:color_basis}.
In this basis, we can perform threshold resummation for jet and 
other cross sections.

To compare to explicit calculations at the two-loop level, however,
and to generalize to higher numbers of external partons,
it is convenient to make contact with a somewhat different 
notation, in which the color interactions of soft gluons is represented
by a color matrix ${\bf T}_i^a$ for the insertion of a gluon on external
line $i$, with ${\bf T}_i^a$ a generator in the color representation 
of that parton $i$, whose color-matrix (rather than generator) 
indices are summed against those of the lower-order amplitude 
that is ``dressed" by this soft gluon.
In the notation of Eq.~(\ref{amp}) above for the color content of
an amplitude, the action of the generators may be made explicit 
as the action of a vector, with indices in the adjoint representation;
for example for $i=1$,
\bea
\label{ampij}
\Bigl[ {\bf T}^a_1\, \left| {\cal M}_\f \right \rangle \Bigr]_{d_1,r_2\dots}
 \equiv
 \M^{[\f]}_{L}\left(\beta_j,\frac{Q^2}{\mu^2},\as(\mu^2),\vep \right)
\,  \delta_i\left({T^a}\right)_{d_1r_1}\, \left(c_L\right)_{\{r_1,r_2\dots \}}
\, ,
\eea
where $\delta_i=\pm 1$ absorbs minus signs associated with 
antiparticles and crossing.  In the convention of Ref.~\cite{catani98},
$\delta_i=1$ when $i$ is an eikonal line representing an outgoing 
quark or gluon, or incoming antiquark; $\delta_i=-1$ for an incoming 
quark or gluon, or outgoing antiquark.  
Defined in this way, the vector color generator matrices obey the
fundamental relation $\sum_i{\bf T}_i=0$, which is an expression
of gauge invariance.   The ${\bf T}_i$s are conventionally
normalized to ${\bf T}_i \cdot {\bf T}_i =C_i$, with $i=q,g$.

A gluon exchanged between two parton lines $i$ and $j$ produces the
product ${\bf T}_i \cdot {\bf T}_j$, which acts on an amplitude in
a  fashion precisely similar to Eq.~(\ref{ampij}).  
This notation allows for a convenient iterative expression for 
color exchange due to soft gluon exchange,
without requiring an explicit choice for the color basis.

The color-space notation above may be applied to the computation of
the soft function as well as to the amplitude itself.   
Consider the soft function at a single loop, determined by 
the one-loop soft anomalous dimension.  As we have seen, 
the latter is built up from the contributions of soft gluon 
exchanges between pairs of eikonal lines.  From each exchange, 
the contribution to the anomalous dimension
is found from Eq.~(\ref{Eq:gamma_oneloop}), where the one-loop
single-pole term in $Z_{S_\f}$ equals the one-loop UV pole
term computed from the corresponding diagram.

Each diagrammatic contribution, then, is proportional to a product
${\bf T}_i\cdot {\bf T}_j$ acting on the lower-order amplitude, multiplied
by the result of the eikonal integral.  
Referring to Fig.~\ref{Fi:one_loop}, the relevant
single-pole coefficients are given in Eq.~(\ref{1loopFs}).
The action of ${\bf \Gamma}_{S_\f}^{(1)}$ on the color tensor
is the sum of all such terms, with a subtraction for the jet anomalous
dimensions; this subtraction is proportional to the identity matrix in 
color space, as in Eq.~(\ref{GammaSIJdef}).  This gives
\bea
{\bf \Gamma}_{S_\f}^{(1)}
\left(\frac{s_{ij}}{\mu^2}\right)|\mathcal{M}_\f \rangle
&=& \left[\,
\frac{1}{2}\sum_{i\in \f}\sum_{j\ne i\in \f}
(\delta_i{\bf T}_i\cdot \delta_j{\bf T}_j)\, 
\left( \frac{-2\vep\, F^{s.p.}_{s_{ij}}}{(\as/\pi)}\right)  
\, - \, \sum_{i\in \f}
\Gamma_2^{[i]\,(1)}\left(\frac{u_0}{v^2}\right)\, \right]
\; |\mathcal{M}_\f
\rangle\nonumber\\
&=&
\left[\, - \frac{1}{2}\sum_{i\in \f}\sum_{j\ne i} {\bf T}_i\cdot {\bf T}_j
  \,
\ln\left(\frac{-s_{ij}}{Q^2v^2}\right)\,
  -\, \frac{1}{2} \sum_{i\in \f}  C_i
  \, \ln\left(\frac{\mu^2}{Q^2v^2}\right)\, \right]\; |\mathcal{M}_\f\rangle
\nonumber\\
&=& 
\frac{1}{2}\sum_{i\in \f}\sum_{j\ne i}{\bf T}_i\cdot {\bf T}_j \,
  \,\ln\left(\frac{\mu^2}{-s_{ij}}\right)|\mathcal{M}_\f\rangle \,,
\label{GammaS1T}
\eea
where $s_{ij} =(p_i+p_j)^2$, with all momenta defined to flow into
(or out of) the amplitude.  For the four-parton case above, 
$s_{12}=s$, $s_{13}=t$,
and so forth.  The overall 1/2 compensates for double counting in the sum.
To derive the final result, we have used 
the explicit forms of the $\delta_i$s described above,
as well as the identities $\sum_i{\bf T}_i=0$ and 
${\bf T}_i\cdot {\bf T}_i=C_i$
(in the quark-scattering case, all $C_i=C_F$).   In this notation
the color identities enforce the cancellation of the
 collinear-sensitive $\ln(1/v^2)$ 
terms.

Identical considerations apply to the two-loop case.  
The nonvanishing anomalous dimension matrix is again a sum of 
diagrammatic contributions, corresponding to gluon exchange 
processes involving two eikonal lines only.  As we have seen, 
these contributions have the same color-generator structure, 
${\bf T}_i\cdot {\bf T}_j$, found at one loop.  
The 3E diagrams have a more complicated color
structure, but they do not contribute to the two-loop
soft anomalous dimension matrix.

To be more specific, we saw that diagram $(h)$ in Fig.~\ref{Fi:2ld} 
can be organized into antisymmetric and symmetric color structures, which
can be represented as commutators and anti-commutators of 
one-loop color structures, of the form
$[{\bf T}_i\cdot {\bf T}_j \, , \, {\bf T}_j\cdot {\bf T}_k]$
and $\{ {\bf T}_i\cdot {\bf T}_j \, , \, {\bf T}_j\cdot {\bf T}_k \}$.
Note that the antisymmetric quantity can be written as,
\begin{equation}
[ {\bf T}_i \cdot {\bf T}_j \, , {\bf T}_j \cdot {\bf T}_k ]
= {\bf T}_i^a\, [{\bf T}_j^a,{\bf T}_j^b]\, {\bf T}_k^b
= - i \, f^{abc} \, {\bf T}_i^a \,{\bf T}_j^b {\bf T}_k^c \,.
\label{colorcommutator} 
\end{equation}
As the final form shows, it is totally antisymmetric under 
permutations of the three eikonal lines.   This is also the
form of the color factor for the other type of 3E diagram,
the three-gluon diagram $(g)$ in Fig.~\ref{Fi:2ld}.

As emphasized above, the velocity factors multiplying
both commutator and anti-commutator structures
vanish. (In the case of the anti-commutator, the vanishing
occurs after adding the one-loop counterterms.)  
Nevertheless, we display the commutator in Eq.~(\ref{colorcommutator}), 
because it has occurred in the literature before. 
We will encounter such terms below in our analysis of
explicit two-loop calculations, and show how they
are consistent with the specific solution for the
soft anomalous dimension, Eq.~(\ref{Sto2}), in
which this combination of color generators does not appear. 

\subsection{Generalizations}

The analysis given above applies far beyond the $2\to 2$
quark-antiquark scattering amplitude.
When the soft anomalous dimension is expressed in terms of 
color generators, as in Eq.~(\ref{GammaS1T}) at one loop, 
and using this equation and Eq.~(\ref{Gamma2complete}) to do
so at two loops, the result is slightly less explicit than, say,
Eq.~(\ref{Eq:1l_gammaS}), but it is much more general.  
When we generalize from quark and antiquark to gluon lines,
and when we add more partons in the final state, 
the only change in our considerations above is
to change the color generators ${\bf T}_i$,
and sum over more variables $i$ and $j$.
The eikonal momentum integrals 
that give rise to
the coefficients of the generators are the same 
for any choice of parton pairs or triplets.

In these terms, the two-loop results organized in 
Eqs.~(\ref{GammaS1T}) and (\ref{Gamma2complete})
are not limited to quark-antiquark scattering, but apply to 
the scattering of any flavor combination.  Furthermore, 
these relations are by no means limited to $2\to 2$ scattering, and apply
to any $2\to n$ process, as in multi-jet production.  These
results, therefore, are a step toward threshold and related
resummations in hadronic scattering~\cite{KOS,BSZ} at the level of 
next-to-next-to-leading logarithm.  

At present, however, for the purposes of
resummation we must still rely upon the explicit form
of the matrix as in Eq.~(\ref{expoS}) to generate the amplitude at
arbitrary orders.  Anticipating further applications,
it will be useful to investigate flexible choices of
color basis, perhaps based on the trace notation
described, for example, in Ref.~\cite{DixonTASI95}
(this point was noted in Ref.~\cite{BSZ}).
We reserve these considerations for future work.

\section{Single poles at NNLO}

In this section, we combine the expansions of the jet
and soft functions in the ``minimal" factorized amplitude,
Eqs.~(\ref{facampmin}) and (\ref{statecalH}), and give 
an explicit expression for
infrared poles to two loops, including single poles.   
We go on to compare these ``postdictions" of the
two-loop single pole terms to the results of explicit calculations,
and verify that they agree.
Traditionally, these results have been presented in a form
proposed some time before by Catani~\cite{catani98},
and we will briefly review this formalism and relate it
to  the two-loop expansion of our resummed expressions.

\subsection{Two-loop poles from the factorized amplitude}

Here, as above, we adopt the notation $f(\as)=\sum_n (\as/\pi)^n f^{(n)}$.
In this notation, we can express the Born and one-loop amplitudes 
for process $\f$ in terms of the factorized jet, soft and hard functions
of Eq.~(\ref{statecalH}) as
\bea
 \left |{\cal M}_\f^{(0)} \right\rangle &=&  \left |{\cal H}_\f^{(0)}   \right\rangle
 \\
\left | {\cal M}_\f^{(1)}\right\rangle
&=&
\left( \sum_{i\in \f} E^{[i](1)} + {\bf S}_\f^{(1)}\, \right) \left |{\cal M}_\f^{(0)} \right\rangle
+
\left |{\cal H}_\f^{(1)} \right\rangle 
\nonumber\\
&=& \left( - \frac{1}{4}\, 
\sum_{i\in \f} \left(\, \frac{1}{2\varepsilon^2}\, \gamma^{[i](1)}_K + \frac{1}{\varepsilon}
 {\cal G}^{[i](1)}_0\, \right)
+ \frac{1}{2\vep}{\bf\Gamma}_{S_\f}^{(1)}\, 
\right) \left |{\cal M}_\f^{(0)} \right\rangle\
+\
\left |{\cal H}_\f^{(1)} \right\rangle \, ,
\label{M1js}
\eea
where for the jet functions we have used the minimal form~(\ref{calJdef}).
In the second equality for $|{\cal M}_\f^{(1)}\rangle$,  we  have used
explicit expressions for the jet functions and the soft matrix, the latter
from Eq.~(\ref{Sto2}).  Using these results, we find for the two-loop amplitude
\bea
\left | {\cal M}_\f^{(2)}\right\rangle
&=&
\left[\, \frac{1}{2} \left( \sum_{i\in \f} E^{[i](1)} + {\bf S}_\f^{(1)}\, \right)^2
+ \sum_{i\in \f} E^{[i](2)} + {\bf S}_\f^{(2)} - \frac{1}{2}\left({\bf S}_\f^{(1)}\right)^2 \, \right]\, \left | {\cal M}_\f^{(0)}\right\rangle
\nonumber\\
&\ & \hspace{10mm} + 
\left( \sum_{i\in \f} E^{[i](1)} + {\bf S}_\f^{(1)}\, \right) \left |{\cal H}_\f^{(1)} \right\rangle\
+\  \left |{\cal H}_\f^{(2)} \right\rangle
\, .
\label{fullM2}
\eea
Now both the $E^{[i](n)}$ and the ${\bf S}_\f^{(n)}$ are given 
by sums of pure poles in $\vep$.
As a result, their squares and products all begin at $1/\vep^2$.  
At two loops, then, the single-pole terms that multiply the 
Born amplitude $|{\cal M}_\f^{(0)}\rangle$ in Eq.~(\ref{fullM2}) 
are given entirely by the single poles in $E^{[i](2)}$ and 
${\bf S}_\f^{(2)}$.  From Eq.~(\ref{calJdef})
for the jets, and Eq.~(\ref{Sto2}) for the soft matrix, these poles
are found from the coefficients of the soft and jet 
anomalous dimensions, 
\bea
 \left | {\cal M}_\f^{(2)}\right\rangle
&=&
 \Bigg[\, 
\frac{1}{2} \left( \sum_{i\in \f} E^{[i](1)} 
+ \frac{1}{2\vep}{\bf\Gamma}_{S_\f}^{(1)}\, \right)^2
+ \sum_{i\in \f} \sum_{j=2}^3 E_j^{[i](2)} 
- \frac{\beta_0}{16\vep^2}\, {\bf \Gamma}_{S_\f}^{(1)}\,
 \Bigg]\, \left | {\cal M}_\f^{(0)}\right\rangle
\nonumber\\
&\ & \hspace{5mm}
+
\left[ \, \sum_{i \in \f}E_1^{[i](2)} + \frac{1}{4\vep}\,
{\bf \Gamma}_{S_\f}^{(2)}\, \right] \, \left | {\cal M}_\f^{(0)}\right\rangle
\nonumber\\
&\ & \hspace{5mm}
+\ \left( \sum_{i\in \f} E^{[i](1)} 
+ \frac{1}{2\vep}{\bf\Gamma}_{S_\f}^{(1)}\, 
 \right) 
 \left |{\cal H}_\f^{(1)} \right\rangle
+ \left |{\cal H}_\f^{(2)} \right\rangle
\, ,
\label{M2expand}
\eea
where we have separated the double- and 
higher-order pole terms from single-pole terms that multiply the 
Born amplitude, followed by single poles times the
one-loop hard scattering, and finally the two-loop hard scattering.

{}From Eq.~(\ref{Enmvalues}), the two-loop single-pole terms that 
multiply the Born amplitude $|{\cal M}_\f^{(0)}\rangle$
in Eq.~(\ref{M2expand}) are given by
\bea
\left[ \, \sum_{i \in \f}E_1^{[i](2)} + \frac{1}{4\vep}\,
{\bf \Gamma}_{S_\f}^{(2)}\, \right] \, \left | {\cal M}_\f^{(0)}\right\rangle
=
\frac{1}{\vep}\, \left[
- \frac{{\cal G}^{[i](2)}_0}{8} 
+ \frac{\beta_0\, {\cal G}^{[i](1)}{}'(0) }{32}
+\frac{K}{8}\,
{\bf \Gamma}_{S_\f}^{(1)}\, \right] \, \left | {\cal M}_\f^{(0)}\right\rangle\, ,
\label{Bornsingleexpand}
\eea
where we recall the notation of Eq.~(\ref{calGnotation}) for the
coefficients ${\cal G}^{[i](n)}(\vep)$.
Given that the one- and two-loop 
${\cal G}^{[i](n)}$ have been known for a long
time, and that we have just calculated 
the two-loop soft anomalous dimension matrix,
this expression provides an explicit form for the
intrinsic two-loop single poles in dimensionally
regulated amplitudes.   

Note that a redefinition of ${\bf S}_\f^{(1)}$ to
include a non-pole term would
both change the definition of $|{\cal H}_\f^{(1)}\rangle$ at 
one loop in Eq.~(\ref{M1js}),
and introduce single-pole terms into the $(E^{[i](1)}+{\bf S}_\f^{(1)})^2$
contribution to the two-loop expression Eq.~(\ref{fullM2}).
As we shall see below, the Born-times-single-pole terms 
remain invariant under this shift only if the shift commutes with the 
${\bf \Gamma}_{S_\f}^{(1)}$.    
We therefore need all of the expansion~(\ref{M2expand}), in order to
make contact with the results of explicit two-loop calculations 
at the single-pole level.
 
\subsection{One- and two-loop amplitudes in Catani's notation}

To compare to existing calculations, we now review
the notation of Ref.~\cite{catani98}, in which they are normally presented.
We first observe that in this notation, amplitudes are organized in 
powers of $(\as/2\pi)$, rather than  $(\as/\pi)$.  We will
distinguish this trivial difference below by
a prime in the color states, as $|\M_\f'{}^{(n)}\rangle
= 2^n |\M_\f{}^{(n)}\rangle$.

In this formalism, the single- and double-pole structure of
one-loop amplitudes is expressed in terms of the 
color generator operators introduced above,
\bea
\I_\f^{(1)}(\vep) = 
\frac{1}{2}\,  \frac{{\rm e}^{-\vep\psi(1)}}{\Gamma(1-\vep)}\, 
\sum_{i\in \f}\sum_{j\ne i}({\bf T}_i\cdot {\bf T}_j)
\, \left[\, \frac{1}{\vep^2}
+ \frac{\gamma_i}{{\bf T}_i^2}\frac{1}{\vep}\, \right]
\, \left(\frac{\mu^2}{-s_{ij}}\right)^\vep\, ,
\label{I1def}
\eea
with $\psi(1) = - \gamma_E$ the logarithmic derivative of the Gamma function.
Here $\gamma_g=\beta_0/2$ and ${\bf  T}_g^2=C_A$ for gluons ($i=g$), 
and $\gamma_i/ {\bf  T}_i^2=3/2$ for $i=q$ or $\bar q$.
The poles of the one-loop amplitude in color-state notation are 
then represented as
\bea
\left| \M'_\f{}^{(1)}\right \rangle =  \I_\f^{(1)}(\vep)\, \left| \M'_\f{}^{(0)}\right \rangle
+ \left| \M'_\f{}^{(1){\rm fin}}\right \rangle\, .
\label{M1I1}
\eea

The explicit relation to the resummation
formalism at one loop is found by expanding $\I^{(1)}_\f$ in powers of $\vep$,
\bea
\I_\f^{(1)} \left| \M'_\f{}^{(0)}\right \rangle
&=&
\Bigg[\, 2\sum_{i\in \f} E^{[i](1)}(\vep) 
+ \frac{1}{\vep}{\bf\Gamma}_{S_\f}^{(1)}
 + \frac{\zeta(2)}{4} \, \sum_{i\in \f} {\bf T}_i^2 \nonumber\\
&\ & \hspace{-5mm} +
\frac{1}{2}\, \sum_{j\ne i}({\bf T}_i\cdot {\bf T}_j)\, \left( \frac{1}{2}\ln^2 \left(\frac{\mu^2}{-s_{ij}}\right)
+ \frac{\gamma_i}{{\bf T}^2_i}\, \ln \left(\frac{\mu^2}{-s_{ij}}\right)\right) + {\cal O}(\vep)
\, \Bigg]\, \left| \M'_\f{}^{(0)}\right \rangle
\nonumber\\
&\equiv& \Bigg[\, 2\sum_{i\in \f} E^{[i](1)}(\vep) 
+ \frac{1}{\vep}{\bf\Gamma}_{S_\f}^{(1)} 
+ \I^{(1){\rm fin}}_\f\, \Bigg]\, \left| \M'_\f{}^{(0)}\right \rangle
\, .
\label{I1expand}
\eea
Taking into account the overall factor of two from the expansion 
in $\as/2\pi$, the pole terms in Eq.~(\ref{M1I1})
are thus identical to those in Eq.~(\ref{M1js}).  The matrix 
$\I_\f^{(1)}$ generates
as well explicit $\mu$-dependent finite contributions
contained in $\I^{(1){\rm fin}}_\f$, which in the ``minimal"
factorization are absorbed into the one-loop hard function 
$|{\cal  H}_\f\rangle$.
The one-loop infrared finite amplitudes are related by
\bea
\left |{\cal  H}'_\f{}^{(1)} \right \rangle 
= \left | {\cal M}'_\f{}^{(1){\rm fin}}\right\rangle
+ \I^{(1){\rm fin}}_\f\, \left | {\cal M}'_\f{}^{(0)}\right\rangle\, .
\label{H1toM1fin}
\eea
This is an example of a finite shift of the sort mentioned above,
which redefines the finite function at one loop.   

At two loops, Ref.~\cite{catani98} predicted the fourth- through
second order poles in terms of the generators $\I_\f^{(1)}$, and
absorbed the then-unknown single pole contributions 
in terms of a color operator ${\bf H}_\f^{(2)}$,
\bea
\I_\f^{(2)}(\vep)
&=&   
- \frac{1}{2}\, \I_\f^{(1)}(\vep)\, 
\left(\,  \I_\f^{(1)}(\vep)\, +\, \frac{\beta_0}{\vep}\right) \, 
\nonumber\\
&\ & \hspace{5mm}
  + \frac{{\rm e}^{\vep\psi(1)}\Gamma(1-2\vep)}{\Gamma(1-\vep)}\, 
  \left (\, \frac{\beta_0}{2\vep} + K  \right )\, \I_\f^{(1)}(2\vep)
 +  {\bf H}_\f^{(2)}(\vep) \, .
\label{I2def}
\eea
The two-loop amplitude is then organized as
\bea
\left | {\cal M}'_\f{}^{(2)}\right\rangle
= \I_\f^{(2)}(\vep)\, \left | {\cal M}'_\f{}^{(0)}\right\rangle + 
 \I_\f^{(1)}(\vep)\, \left | {\cal M}'_\f{}^{(1)}\right\rangle
 + \left | {\cal M}'_\f{}^{(2){\rm fin}}\right\rangle\, .
\label{M2I2}
\eea
In the intervening years, the color generators ${\bf H}^{(2)}_\f$ 
have been determined by
matching to the single-pole structure of explicit
two-loop QCD scattering amplitude calculations, for example
$gg\to gg$, $q\bar q\to gg$, $q\bar{q} \to q\bar{q}$, 
and $e^+e^-\to q \bar{q}g$~\cite{Twoloopqqqqqqgg,%
Twoloopgggg,Twoloopgggghel,ee3partonNNLO,BDK04}.
Here we follow Ref.~\cite{BDK04} and write
\bea
{\bf H}_\f^{(2)}(\vep)
=
\frac{1}{4\vep}\, \left \{
\, \sum_{i\in \f} H^{(2)}_i  
+ \hat{\bf H}^{(2)}_\f \, \right \}\ +\ {\cal  O}(\vep)\, ,
\label{HH} 
\eea
where we split the single-pole
factor into a color-diagonal term, which can be represented
as a sum of constants $H_i^{(2)}$ for each external parton $i$, 
and a matrix $\hat{\bf H}_\f^{(2)}$ that includes all color mixing.
This matrix can be written as
\cite{Twoloopqqqqqqgg,%
Twoloopgggg,Twoloopgggghel,ee3partonNNLO,BDK04}
\bea
\hat{\bf H}_\f^{(2)}=
i\, \sum_{(i, j, k)}\, f_{a_1a_2a_3}\,  
{\bf T}_i^{a_1}\,  {\bf T}_j^{a_2}\, 
 {\bf T}_k^{a_3}\, 
\ln\left(\frac{-s_{ij}}{-s_{jk}}\right)\,
\ln\left(\frac{-s_{jk}}{-s_{ki}}\right)\,
\ln\left(\frac{-s_{ki}}{-s_{ij}}\right)\, ,
\label{H2def}
\eea
where the sum is over distinguishable
but unordered triplets of 
external lines $(i,j,k)$.  We note the similarity to
the color structure from the 3E diagrams, Eq.~(\ref{colorcommutator}).
We emphasize that this form has been 
obtained directly only for processes with at most four partonic legs.
In Ref.~\cite{BDK04} it was also shown to be consistent with the 
proper collinear behavior of the $2\to n$ gluon amplitudes.

It is these expressions that we
will compare to the two-loop expansion of
the factorized amplitude, Eqs.~(\ref{M2expand})
and ~(\ref{Bornsingleexpand}).
Rather than provide explicit expressions at this point for the
constants $H_i^{(2)}$
from~\cite{BDK04}, we will derive below expressions 
relating the constants $H_i^{(2)}$ to the jet anomalous dimensions
(in $\overline{\rm MS}$ scheme).   
Here we will find useful an identity found in Ref.~\cite{RSvN04}.
The matrix $\hat{\bf H}_\f^{(2)}$ in Eq.~(\ref{H2def}) will emerge from
our results for the two-loop soft anomalous dimension matrix,  
plus the effects of a one-loop finite color-mixing term.
We now turn to this exercise.

\subsection{${\bf H}^{(2)}$ from the anomalous dimensions}

Inserting the definition of $\I_\f^{(2)}$, Eq.~(\ref{I2def}),
into Eq.~(\ref{M2I2}) and expanding to the accuracy of $\vep^0$, 
we readily find
\bea
\left| \M'_\f{}^{(2)}\right \rangle
&=&  \frac{1}{2}\, 
\left(\,  \I_\f^{(1)}(\vep)\, \right)^2 \, \left| \M'_\f{}^{(0)}\right\rangle
\nonumber\\
&\ &
\hspace{-10mm}
  + \left [\,   \frac{\beta_0}{2\vep}\,  \left(\, \I_\f^{(1)}(2\vep) - \I_\f^{(1)}(\vep) \right)
   + \left (\, K +   \frac{3\vep\zeta(2)}{4}\,\beta_0\, \right )
\, \I_\f^{(1)}(2\vep) +  {\bf H}_\f^{(2)}(\vep) 
+ {\cal O}(\vep^0)\, \right]\, \left| \M'_\f{}^{(0)} \right \rangle
\nonumber\\
&\ &\hspace{-10mm}
 +\ \I_\f^{(1)}(\vep)\, \left| \M'_\f{}^{(1){\rm fin}} \right \rangle 
\, + \,  \left| \M'_\f{}^{(2){\rm fin}}  \right \rangle \, .
\label{Mprimestart}
\eea
We will relate this expression to the single-pole
result from the factorized amplitude, Eqs.~(\ref{M2expand}) 
and (\ref{Bornsingleexpand}).

The single-pole terms in Eq.~(\ref{Mprimestart}) that multiply the 
Born amplitude come from two sources: 
the $(\I_\f^{(1)})^2$ operator on the first line,
and the terms in the square brackets on the second line, 
which also include finite corrections indicated by $+ \, {\cal O}(\vep^0)$. 

To make contact with the expansion of the resummed
amplitude, Eq.~(\ref{M2expand}), we first separate the poles
in each of the $\I^{(1)}_\f$ terms, according to Eq.~(\ref{I1expand}), 
\bea
\left| \M'_\f{}^{(2)}\right \rangle
&=& 
\left [\, \frac{1}{2}\,  \left(\,  2\sum_{i\in \f} E^{[i](1)}(\vep)  
+ \frac{1}{\vep}{\bf\Gamma}_{S_\f}^{(1)} \, \right)^2
+ 2\left(\,  \sum_{i\in \f} E^{[i](1)}(\vep)  \, \right)\ 
\I^{(1){\rm fin}}_\f  \, \right]\, \left| \M'_\f{}^{(0)}\right \rangle
\nonumber\\
&\ & 
+ \, \frac{1}{2}\, \left[\, \left( \I^{(1){\rm fin}}_\f 
\, \frac{1}{\vep}{\bf\Gamma}_{S_\f}^{(1)}  
   + \frac{1}{\vep}{\bf\Gamma}_{S_\f}^{(1)} \, \I^{(1){\rm fin}}_\f \, \right)
+  \left(\I_\f^{(1){\rm fin}}\right)^2\, \right]\, \left| \M'_\f{}^{(0)} \right \rangle 
\nonumber\\
&\ &
  + \Bigg [\,   \frac{\beta_0}{2\vep}\, 
   \left(\,  2\sum_{i\in \f} E^{[i](1)}(2\vep)  - 2\sum_{i\in \f} E^{[i](1)}(\vep)
-  \frac{1}{2\vep}{\bf\Gamma}_{S_\f}^{(1)} \right)
+ \frac{3\zeta(2)\beta_0}{8\vep} \sum_{i\in \f}E_2^{[i](1)}
  \nonumber\\
  &\ &
  \hspace{20mm}
  + \ K
\, \left(\, \,  2\sum_{i\in \f} E^{[i](1)}(2\vep)  + \frac{1}{2\vep}{\bf\Gamma}_{S_\f}^{(1)}
\, \right) 
+  {\bf H}_\f^{(2)}(\vep)
+ \I_\f^{(2){\rm fin}}\, \Bigg]\, \left| \M'_\f{}^{(0)} \right \rangle
\nonumber\\
&\ & 
  \hspace{0mm}
+ \Bigg[\, 2\sum_{i\in \f} E^{[i](1)}(\vep) 
+ \frac{1}{\vep}{\bf\Gamma}_{S_\f}^{(1)} 
+ \I^{(1){\rm fin}}_\f\, \Bigg]\, \, \left| \M'_\f{}^{(1){\rm fin}} \right \rangle 
\ +\  \left| \M'_\f{}^{(2){\rm fin}}  \right \rangle 
\, ,
\eea
where in $\I^{(2){\rm fin}}_\f$ we isolate the finite terms from $\I^{(2)}_\f$ 
that multiply the Born amplitude. 
Comparison with Eq.~(\ref{M2expand}) requires further that we
commute the soft anomalous dimension matrices with poles to the left of the
finite amplitudes, and that we also re-express $|\M'{}^{(1){\rm fin}}\rangle$
in terms of $|{\cal H}'{}^{(1)}\rangle$ using Eq.~(\ref{H1toM1fin}).  
The first step, in particular, leads to an
additional commutator contribution at the level of the single poles
times the Born amplitude,
\bea
\left| \M'_\f{}^{(2)}\right \rangle
&=&  \frac{1}{2}\, 
\left(\,  2\sum_{i\in \f} E^{[i](1)}(\vep)  
+ \frac{1}{\vep}{\bf\Gamma}_{S_\f}^{(1)} \, \right)^2
\, \left| \M'_\f{}^{(0)}\right \rangle
\nonumber\\
&\ &
 \hspace{5mm}
  + \left [\,   \frac{\beta_0}{2\vep}\, 
   \left(\,  2\sum_{i\in \f} E^{[i](1)}(2\vep)  
- 2\sum_{i\in \f} E^{[i](1)}(\vep)
-  \frac{1}{2\vep}{\bf\Gamma}_{S_\f}^{(1)} \right)
 +  K\, \sum_{i\in \f} \frac{E_2^{[i](1)}}{2\vep^2}\,
\right ]\, \left| \M'_\f{}^{(0)} \right \rangle
  \nonumber\\
  &\ &
  \hspace{5mm}
+ \Bigg[\, \frac{3\zeta(2)\beta_0}{8\vep} \sum_{i\in \f}E_2^{[i](1)}
   + K
\, \left(\, \,  \sum_{i\in \f} \frac{E_1^{[i](1)}}{\vep} 
 + \frac{1}{2\vep}{\bf\Gamma}_{S_\f}^{(1)}
\, \right) 
\nonumber\\
&\ & 
  \hspace{35mm}
+ \, \frac{1}{2\vep} \left [   
\I^{(1){\rm fin}}_\f \, ,\, {\bf\Gamma}_{S_\f}^{(1)}    \right ]
+  {\bf H}_\f^{(2)}(\vep)\, \Bigg]\, \left| \M'_\f{}^{(0)} \right \rangle
\nonumber\\
&\ & 
  \hspace{5mm}
+ \Bigg[\, 2\sum_{i\in \f} E^{[i](1)}(\vep) 
+ \frac{1}{\vep}{\bf\Gamma}_{S_\f}^{(1)} 
\, \Bigg]\, \, \left| {\cal H}_{\f}'{}^{(1)} \right \rangle 
\ +\  \left| {\cal H}_{\f}'{}^{(2)} \right \rangle\,.
\label{M2I2expand}
\eea
Here we have organized the expression just as in Eq.~(\ref{M2expand}),
starting  with the square of one-loop pole terms, two-loop
second- and third-order poles, and then first-order poles, all
times the Born amplitude,  followed by poles times the one-loop
hard amplitude and the finally the two-loop hard part,
\bea
\left| {\cal H}'{}^{(2)} \right \rangle \equiv
\left( \I_\f^{(2){\rm fin}} - \frac{1}{2}\left(\I^{(1){\rm fin}}_\f\right)^2\right)\, \left| \M'_\f{}^{(0)} \right \rangle
+
\I^{(1){\rm fin}}_\f\,  \left| {\cal H}_{\f}'{}^{(1)} \right \rangle 
+ \left| \M'_\f{}^{(2){\rm fin}}  \right \rangle\, .
\eea
We are now ready to compare this expression to the
two-loop single pole terms of Eq.~(\ref{Bornsingleexpand}).
Higher-order poles can easily be checked in a similar manner~\cite{TYS}.

Consider first the matrix parts of Eqs.~(\ref{Bornsingleexpand}) 
and (\ref{M2I2expand}).
Recalling the factor of 4 associated with changing from
the coefficient of $(\as/\pi)^2$ to $(\as/2\pi)^2$, we see
that the $K\, {\bf\Gamma}_{S_\f}^{(1)}$ term is identical
in the two expressions.  Consistency then requires
the remarkable result that the commutator of $\I_{\f}^{(1){\rm fin}}$ 
in Eq.~(\ref{I1expand}) with the one-loop soft anomalous dimension 
in Eq.~(\ref{GammaS1T}) precisely cancel the two-loop 
$\hat {\bf H}_\f^{(2)}$ function as defined in Eq.~(\ref{H2def}),
\bea
 \left [   \I^{(1){\rm fin}}_\f \, ,\, {\bf\Gamma}_{S_\f}^{(1)}    \right ]
= -\ \frac{1}{2} \, \hat{\bf H}^{(2)}_\f \, .
 \label{commutator}
\eea
In fact, 
a compact calculation, given in
Appendix \ref{comapp}, shows that Eq.~(\ref{commutator}) indeed
holds for the explicit matrix $\hat{\bf H}_\f^{(2)}$ given
in Eq.~(\ref{H2def}), for arbitrary $2\to 2$ processes,
and also for $2\to n$ processes where all particles are identical.
For those processes with five or more partons for which the quantities
$\gamma_i/{\bf T}_i^2$ are not all identical, the commutator is
more complicated, as can be seen by inspecting Eq.~(\ref{I1expand}),
and as discussed in Appendix \ref{comapp}.  Because we
know the anomalous dimension matrix for all these processes,
however, Eq.~(\ref{commutator}) can be turned around
and taken as a definition of the corresponding matrices $\hat{\bf H}_\f^{(2)}$.
Recently, the soft anomalous dimension matrix $\Gamma_{S_\f}^{(2)}$ 
for $2\to 2$ processes was computed~\cite{JKPS} by making use of 
just this connection to $\hat{\bf H}_\f^{(2)}$, along with the 
explicit results for $\hat{\bf H}_\f^{(2)}$ for quark-quark 
scattering~\cite{Twoloopqqqqqqgg}.

The remaining, color-diagonal, single-pole terms in 
Eq.~(\ref{M2I2expand}) are found using the values of the one-loop
quantities $E^{[i](1)}$ given in Eq.~(\ref{Enmvalues}), 
and the form of ${\bf H}_\f^{(2)}$ in given in Eq.~(\ref{HH}).  
Then the single-poles times Born amplitudes
of Eq.~(\ref{M2I2expand}) are given by
\bea
\left| \M_\f'{}^{(2)}\right \rangle_{\rm single\ pole\, \times \, Born}
&=&  \frac{1}{\vep}\;\sum_{i \in \f}  \left [  \frac{H_i^{(2)}}{4} 
- \frac{3\zeta(2)\beta_0}{32}\, C_i
- \frac{K\,  \G_0^{[i]\,(1)}}{4} 
\,   \right] \, \left| \M'_\f{}^{(0)}\right \rangle
+ \, \frac{K}{2\vep}\;
 {\bf \Gamma}_{S_\f}^{(1)}
\, \left| \M'_\f{}^{(0)}\right \rangle\, ,
\nonumber\\
\label{M2singleB1}
\eea
where we have suppressed dependence that contributes 
only at the level $\vep^0$.
The  comparison of Eq.~(\ref{M2singleB1}) with the expansion from the
resummed amplitude, Eq.~(\ref{Bornsingleexpand}), is now trivial.  
We simply appeal to the striking identity noted explicitly by 
Ravindran, Smith and van Neerven~\cite{RSvN04},
which in our notation is written as
\bea
 \frac{H_i^{(2)}}{4} 
- \frac{3\zeta(2)\beta_0}{32}\, C_i
- \frac{K\, \G_0^{[i]\,(1)}}{4} 
\ =\ 
4 E_1^{[i]\,(2)}\,,
\label{HitoE}
\eea
where $E_1^{[i]\,(2)}$ is given in Eq.~(\ref{Enmvalues}).
In Ref.~\cite{RSvN04} this expression was observed to imply
a close relationship  between the $H_i^{(2)}$  constants and
the form factors.  We now see that,  aside from color
mixing, all the single-pole terms are identical
to those in the form factors.
Indeed, the precise terms relating the $H_i^{(2)}$ to the
single-pole residues of the elastic form factor are present simply to
cancel a set of ``extra" single-pole terms generated from
the expansion of $\I_\f^{(1)}$ in the two-loop amplitude.
As in the case of the color-mixing anomalous dimensions,
we can also consider Eq.~(\ref{HitoE}) as a {\it definition}
of the constants $H^{(2)}_i$.

In summary, we have shown that the full single-pole structure of the
two-loop amplitudes can be reconstructed from the same anomalous dimensions
that determine the next-to-next-to-leading poles of the factorized 
jet and soft functions at all orders in pertubation theory.
This relation, and the explicit forms of the anomalous dimensions,
hold for partonic scattering amplitudes with arbitrary numbers 
of external lines.

\section{Conclusions}

We have extended the factorization and resummation formalisms
for exclusive amplitudes in QCD to next-to-next-to-leading 
poles in these amplitudes.  The same anomalous dimension
matrices, calculated here directly for the first time at two loops, control
a variety of resummed cross sections at NNLL.  
These calculations
generalize the determination of the Sudakov anomalous
dimensions to nontrivial color mixing.  

We verified the formalism and anomalous dimensions by showing 
that they allow us to reproduce the very nontrivial color and 
momentum structure of single infrared poles at 
next-to-next-to-leading order for $2\to 2$ processes in the literature.

The calculation of the NNLO soft anomalous dimensions opens the door to 
threshold resummation at next-to-next-to-leading logarithm
for multijet cross sections~\cite{KOSjet,BSZ,Boncianietal}.
Perhaps our most striking result is the discovery that the 
two-loop soft anomalous dimension matrix is obtained from the 
one-loop matrix simply by multiplying by $K\alpha_s/(2\pi)$, 
where $K$ is the constant given in Eq.~(\ref{KDef}).  
This is exactly the same property obeyed by the scalar Sudakov or 
``cusp" anomalous dimension.
Aside from its intrinsic interest, this relation will make
possible next-to-next-to-leading logarithmic resummation formulas
in a closed form, since it will be possible to diagonalize
the two-loop anomalous dimension matrix
independently of the running of the coupling~\cite{webs,CMW,Berger:2002sv}, 
using the same color eigenvectors
found at one loop~\cite{KOSjet,BSZ,Boncianietal,DM05,kyrieleis05}.

Our analysis applies not only to $2\rightarrow n$ processes relevant to
hadronic colliders.  In addition, it applies to the inelastic 
scattering of a parton by a color-singlet source, as in deep-inelastic
scattering, and to the creation of arbitrary
numbers of partons from a color-singlet source in leptonic annihilation.
It will clearly also be of interest to extend this analysis to 
massive external lines.

\subsection*{Acknowledgements}

This work was supported in part
by the National Science Foundation, grants PHY-0098527 and PHY-0354776,
and by the Department of Energy under contract DE--AC02--76SF00515.
We wish to thank Babis Anastasiou, Carola Berger, 
Zvi Bern, Stefano Catani, Yuri Dokshitzer, Nigel Glover, 
David Kosower, Hans K\"uhn, Alexander Penin, Jack Smith 
and Werner Vogelsang for very helpful conversations.   
LD wishes to thank the Kavli Institute for Theoretical
Physics and the Aspen Center for Physics for support during 
a portion of this work,
and GS thanks the Stanford Linear Accelerator Center for hospitality.
The figures were generated using Jaxodraw~\cite{Jaxo}, based
on Axodraw~\cite{Axo}.


\begin{appendix}

\section{Anomalous dimensions}
\label{AnomDimAppendix}

In this appendix, we provide the low-order anomalous dimensions
entering the jet function, as defined in Eqs.~(\ref{jsudakov})
and (\ref{logJmin}).  We give the $n$th-order
coefficients $\gamma_K^{[i]\,(n)}$, $\K^{[i]\,(n)}$ 
and ${\cal G}^{[i]\,(n)}$ in an expansion in powers 
of $\alpha_s(\mu^2)/\pi$,
\bea
\gamma_K^{[i]\,(1)} &=& 2 C_i\,,
\nonumber\\
\gamma_K^{[i]\,(2)} &=& C_i K = C_i \left [ 
C_A \biggl( {67\over18} - \zeta(2) \biggr) - {10\over9} T_F n_F \right] \,,
\nonumber\\
\K^{[i]\,(1)} &=& \frac{1}{2\vep}\gamma_K^{[i]\,(1)} \,,
\nonumber\\
\K^{[i]\,(2)} &=& \frac{1}{4\vep}\gamma_K^{[i]\,(2)}
- \frac{\beta_0}{16\vep^2}\gamma_K^{[i]\,(1)} \,,
\nonumber\\
{\cal G}^{[q]\,(1)} &=& \frac{3}{2}C_F 
+ \vep \frac{C_F}{2}\, \left(  8 - \zeta(2) \right) + {\cal O}(\vep^2) \,,
\nonumber \\
{\cal G}^{[q]\,(2)}_0 &=&
3 C_F^2\, \left[ \frac{1}{16} - \frac{1}{2}\zeta(2) +\zeta(3)\right]
+ \frac{1}{4} C_A C_F 
\, \left [ \frac{2545}{108} + \frac{11}{3}\zeta(2) - 13\zeta(3)\right] 
\nonumber\\
&\ & \hspace{0.1mm} 
- \, C_F T_F n_F \, \left[ \frac{209}{108} + \frac{1}{3}\zeta(2)\right] \,,
\nonumber \\
{\cal G}^{[g]\,(1)} &=& \frac{\beta_0}{2} \,
   - \, \vep \frac{C_A}{2}\, \zeta(2)
                              + {\cal O}(\vep^2) \,,
\nonumber \\
{\cal G}^{[g]\,(2)}_0 &=& 
C_A^2\, \left[ \frac{10}{27} - \frac{11}{12} \zeta(2)
                  - \frac{1}{4} \zeta(3) \right]
+ C_A T_F n_F \left[ \frac{13}{27} + \frac{1}{3} \zeta(2) \right]
+ \frac{1}{2} C_F T_F n_F
\nonumber\\
&&\hskip0mm
 + \, \frac{\beta_1}{4}
  \,,
\eea
where $C_q = C_F$, $C_g = C_A$, and
\be
\beta_1 = \frac{34}{3} C_A^2 
- 4 C_F T_Fn_F - \frac{20}{3} C_A T_F n_F \,.
\label{beta1def}
\ee
The results for ${\cal G}^{[i]\,(n)}$ were obtained from
Ref.~\cite{MVVformfactor}, which also contains results through
three loops.  We shift the gluonic expressions by terms proportional
to $\beta$-function coefficients, which take into account the effects
of the renormalizing the operator $G_{\mu\nu}^a G^{a\,\mu\nu}$, 
as explained in Ref.~\cite{RSvN04}.
Because we only quote results through two-loop order, some of the
results for ${\cal G}^{[i]\,(n)}$ could also have been extracted 
from the two-loop quark electromagnetic form factor~\cite{EMformfactor} 
and from the $gg \to $ Higgs boson amplitude~\cite{Harlander00}.

\section{One-loop velocity factors}
\label{OneloopVelFactAppendix}

\subsection{Basic integrals}

Consider the one-loop $t$-channel diagram shown in 
Fig.~\ref{Fi:one_loop}$(a)$. The velocity factor is given by
\begin{eqnarray}
\label{Eq:1lamp}
F_{t}&=&\left(ig\mu^{\vep}\right)^2\,(v_1\cdot v_3)
\,\int_0^{\infty}d\alpha\int_0^{\infty}d\beta
\,D(v_3\alpha+v_1\beta)\nonumber\\
&=&\left(ig\mu^{\vep}\right)^2\frac{1}{4\pi^{2-\vep}}
\Gamma(1-\vep)\,(v_1\cdot v_3)
\,\int_0^{\infty}d\alpha\int_0^{\infty}d\beta
\,\frac{1}{\left[(v_3\alpha+v_1\beta)^2\right]^{1-\vep}}\,.
\end{eqnarray}
We will use the following change of variables
\begin{equation}
\alpha+\beta\equiv\eta\,, \qquad  \alpha\equiv z\,\eta\,,
\end{equation}
with Jacobian $\eta$. For IR regularization we impose
$\alpha<\frac{1}{\lambda}\iff\eta <\frac{1}{\lambda z}$. 
\footnote{For these one-loop diagrams there is one overall 
IR divergence since all the collinear singularities factorize. 
Therefore it is sufficient to restrict only one of the gluon 
attachments.} Also note that $0<z<1$. 
In terms of the new variables we have
\begin{eqnarray}
F_{t}&=&\left(ig\mu^{\vep}\right)^2\frac{1}{4\pi^{2-\vep}}
\Gamma(1-\vep)\,(v_1\cdot v_3)\,\int_0^1dz
\int_0^{\frac{1}{\lambda z}}d\eta\,\eta^{2\vep-1}
\,\frac{1}{\left[(v_3\,z+v_1\,(1-z))^2\right]^{1-\vep}}\nonumber\\
&=&\left(ig\mu^{\vep}\right)^2\frac{1}{4\pi^{2-\vep}}\Gamma(1-\vep)
\,(v_1\cdot v_3)\,\frac{1}{2\vep}\,\frac{1}{\lambda^{2\vep}}
\int_0^{\infty}dz^{\prime}
\,\frac{1}{\left[(v_3+v_1\,z^{\prime}))^2\right]^{1-\vep}}\,,
\label{FtNewVariables}
\end{eqnarray}
with $z^{\prime}\equiv \frac{1}{z}-1$. From the expansion of the above
expression, the single-pole term and the finite part of the 
one-loop diagram are given by
\begin{eqnarray}
F_{t}&=& - \left(\frac{\alpha_s}{\pi}\right)(v_1\cdot v_3)\,\Bigg\{\frac{1}{2\vep}\,I_1(v_1,v_3)+\frac{1}{2}\,I_m(v_1,v_3)+\frac{1}{2}\left[\ln\left(\frac{\mu^2}{\lambda^2}\right)+\ln(\pi\,\rm{e}^{\gamma_e})\right]\,I_1(v_1,v_3)\Bigg\}\,,\nonumber\\
\label{FtI1Im}
\end{eqnarray}
where we have defined the following integrals
\begin{eqnarray}
I_1(v_1,v_3)&\equiv&\int_0^1dz\frac{1}{(v_3\,z+v_1\,\bar{z})^2}
=\int_0^{\infty}dz^{\prime}\frac{1}{(v_3+v_1\,z^{\prime})^2}\,,\nonumber\\
I_m(v_1,v_3)&\equiv&\int_0^{\infty}dz^{\prime} \, 
\frac{\ln(v_3+v_1\,z^{\prime})^2}{(v_3+v_1\,z^{\prime})^2} \,,
\label{I1Imdef}
\end{eqnarray}
with $\bar{z}\equiv 1-z$.

\subsection{Evaluation of $I_1$ and $I_m$}
\label{IEvalSubAppendix}

We are evaluating eikonal diagrams derived from external Wilson lines. By looking at the usual momentum-space expressions for the amplitudes one can easily see that all these diagrams are scale independent in the eikonal velocities $v_i$. With this property in mind we can simplify the evaluations of the integrals by choosing $v_i^2=1$ without loss of generality.

In order to evaluate $I_1(v_1,v_3)$ we use the following change of
variable~\cite{KR87}
\begin{equation}
e^{2\psi}\equiv\frac{\sqrt{v_3^2}z+\sqrt{v_1^2}\bar{z}\rm{e}^{\gamma_{13}}}
{\sqrt{v_3^2}z+\sqrt{v_1^2}\bar{z}\rm{e}^{-\gamma_{13}}}\, ,
\end{equation}
which gives
\begin{equation}
\frac{d\psi}{dz}= - \frac{1}{2}
\frac{\sqrt{v_1^2v_3^2}(\rm{e}^{\gamma_{13}}-\rm{e}^{-\gamma_{13}})}
{(v_3\,z+v_1\,\bar{z})^2}\,.
\end{equation}
From this change of variable it is very easy to see that
\begin{equation}
\int_0^{\gamma_{13}}d\psi
=\sqrt{v_1^2v_3^2}\,\sinh \gamma_{13}\int_0^1dz
\frac{1}{(v_3\,z+v_1\,\bar{z})^2}\,.
\end{equation}
Therefore we get
\begin{equation}
\label{Eq:I1}
I_1(v_1,v_3)=\frac{1}{\sqrt{v_1^2v_3^2}\,\sinh \gamma_{13}}\,\gamma_{13}\,.
\end{equation}
Note that
\begin{equation}
I_1(v_1,-v_3)=\frac{1}{\sqrt{v_1^2v_3^2}\,\sinh \gamma_{13}}\,(i\pi-\gamma_{13})\,,
\end{equation}
by analytic continuation.

With $v_i^2=1$, $I_m$ can be written as
\begin{eqnarray}
I_m(v_1,v_3)&=&\int_0^{\infty}dy\frac{1}{(y+e^{-\gamma_{13}})(y+e^{\gamma_{13}})}
\ln\left[(y+e^{-\gamma_{13}})(y+e^{\gamma_{13}})\right]\nonumber\\
&=&\frac{1}{2\sinh \gamma_{13}} \int_0^{\infty} dy \Bigg\{
\frac{\ln\left[(y+e^{-\gamma_{13}})(y+e^{\gamma_{13}})\right]}{y+e^{-\gamma_{13}}}
- \frac{\ln\left[(y+e^{-\gamma_{13}})(y+e^{\gamma_{13}})\right]}{y+e^{\gamma_{13}}} 
\Bigg\} \nonumber\\
&=&\frac{1}{2\sinh \gamma_{13}}
\int_0^{\infty}dy \Biggl\{ 
\frac{\ln(y+e^{-\gamma_{13}})}{y+e^{-\gamma_{13}}}
+ \frac{\ln(y+e^{\gamma_{13}})}{y+e^{-\gamma_{13}}}\nonumber\\
&\ &\hspace{32mm} 
- \frac{\ln(y+e^{-\gamma_{13}})}{y+e^{\gamma_{13}}}
- \frac{\ln(y+e^{\gamma_{13}})}{y+e^{\gamma_{13}}} \Biggl\} \,.
\end{eqnarray}
It is easy to verify that the first and last terms in the right-hand side
of the final expression cancel, for example by
changing variables to
$u=y+e^{-\gamma_{13}}$ in the first term and
$u=y+e^{\gamma_{13}}$ in the last.
This leaves us with
\begin{equation}
I_m(v_1,v_3)=\frac{1}{2\sinh \gamma_{13}}
\int_0^{\infty}dy\Bigg\{
\frac{\ln(y+e^{\gamma_{13}})}{y+e^{-\gamma_{13}}}
- \frac{\ln(y+e^{-\gamma_{13}})}{y+e^{\gamma_{13}}}\Bigg\}\,.
\end{equation}
In the high-energy limit 
$\gamma \gg 1$, where $e^{\gamma} \gg e^{-\gamma}$, one 
easily finds
\begin{equation}
\label{Eq:Im}
I_m(v_1,v_3)=\frac{1}{2\sinh \gamma_{13}}
\left[-{\rm Li}_2(-e^{2\gamma_{13}}) + {\rm Li}_2(1) 
+ \mathcal{O}(e^{-\gamma_{13}})\right]
=\frac{1}{\sinh \gamma_{13}}\left[\frac{\pi^2}{6}+\gamma_{13}^2
+\mathcal{O}(e^{-\gamma_{13}}) \right] \,.
\end{equation}
Following the same steps one also gets
\begin{equation}
\label{Eq:Im2}
I_m(v_1,-v_3)=- \frac{1}{\sinh \gamma_{13}}
\left[\frac{\pi^2}{6}+\gamma_{13}^2
+\mathcal{O}(e^{-\gamma_{13}})\right]\,.
\end{equation}

\section{Velocity factors for 2E diagrams}
\label{VelFact2ESubAppendix}

We begin our analysis of the 2E diagrams with the diagram that 
has a three-gluon vertex, Fig.~\ref{Fi:2ld}$(f)$. 
We follow Refs.~\cite{CTFFT,KR87} and write the three-gluon vertex as
\begin{equation}
V_{\mu\nu\rho}(k,l,-k-l)=\bar{V}_{\mu\nu\rho}(k,l)+D_{\mu\nu\rho}(k,l)\,,
\end{equation}
where
\begin{eqnarray}
\bar{V}_{\mu\nu\rho}(k,l)&=&
(2l+k)_{\mu}g_{\nu\rho}+2k_{\rho}g_{\mu\nu}-2k_{\nu}g_{\mu\rho} \,,
\nonumber\\
D_{\mu\nu\rho}(k,l)&=&-l_{\nu}g_{\mu\rho}-(k+l)_{\rho}g_{\mu\nu}\,,
\end{eqnarray}
and where $k$ and $l$ are the loop momenta. 
Indices $\nu$ and $\rho$ attach to the $v_3$ line.
The diagram resulting from $\bar{V}_{\mu\nu\rho}$ is proportional
to $v_3^2$ before integration.  We also note that
the contributions of diagrams of Fig.~\ref{Fi:2ld}$(d)$
and Fig.~\ref{Fi:2ld}$(e)$ are entirely proportional to $v_3^2$
before integration,
since a single gluon propagator attaches twice to the same eikonal line.
These $v_3^2$ contributions turn out to cancel each other in the 
high-energy limit.
We give the result for the $\bar{V}_{\mu\nu\rho}$ contribution below.

The contribution resulting from the $D_{\mu\nu\rho}$ piece 
for the diagram of Fig.~\ref{Fi:2ld}$(f)$ is given by,
\begin{eqnarray}
W_{2E,3g-D}(v_1,v_3) &=& 
- (g\mu^{\vep})^4 \, d^{[t]}_{JI} \, \frac{C_A}{2}
\int\frac{d^Dk}{(2\pi)^D}\frac{d^Dl}{(2\pi)^D}
\frac{1}{k^2}\frac{1}{l^2}\frac{1}{(k+l)^2}
\nonumber\\
&&\hskip30mm \times
\left[\frac{2\,v_3\cdot v_1}{(v_1\cdot k)(v_3\cdot k)}
     +\frac{v_3\cdot v_1}{(v_1\cdot k)(v_3\cdot l)}\right] \,,
\end{eqnarray}
where we have suppressed factors of $i\epsilon$ in the denominators. 
One can evaluate the above expression in either momentum or 
configuration space. We will not review the derivation of
the following result~\cite{KR87}, 
\begin{equation}
\label{Eq:2e3g}
W_{2E,3g-D}^{s.p.}=\left(\frac{\alpha_s}{\pi}\right)^2
\, d^{[t]}_{JI} \, 
\frac{C_A}{2} \Bigg\{\gamma_{13}\coth\gamma_{13}
\left(-\frac{1}{4\vep}+\frac{1}{16\vep}\zeta(2)\right)
+\frac{1}{8\vep}I_2(\gamma_{13})\Bigg\}\,,
\end{equation}
where 
\begin{equation}
I_2(\gamma_{13})\equiv \sinh 2\gamma_{13}\,\int_0^{\gamma_{13}}d\psi \frac{\psi \coth \psi}{\sinh ^2 \gamma_{13}-\sinh ^2 \psi}\ln\left(\frac{\sinh \gamma_{13}}{\sinh \psi}\right)\,.
\end{equation}
We analyze $I_2$ in order to get the high-energy behavior of this
amplitude. We start by writing $I_2$ as
\begin{equation}
\label{Eq:I2_rewritten}
I_2=I_{cth-1}+I_1\,,
\end{equation}
where
\begin{eqnarray}
I_{cth-1}&=&\int_0^{\gamma_{13}}d\psi
\,\left[\frac{\sinh 2\gamma_{13}}{\sinh^2 \gamma_{13}-\sinh^2 \psi}
\ln\left(\frac{\sinh \gamma_{13}}{\sinh \psi}\right)\right]
\times\psi \, (\coth\psi-1)\nonumber\\
&=& 2 \int_0^{\gamma_{13}}d\psi\,\left[
\frac{\sinh 2\gamma_{13}}{\sinh^2 \gamma_{13}-\sinh^2 \psi}
\ln\left(\frac{\sinh \gamma_{13}}{\sinh \psi}\right)\right]
\times\psi\frac{e^{-2\psi}}{1-e^{-2\psi}}\,,
\end{eqnarray}
and where 
\begin{eqnarray}
I_1&=&\int_0^{\gamma_{13}}d\psi
\,\left[\frac{\sinh 2\gamma_{13}}{\sinh^2 \gamma_{13}-\sinh^2 \psi}
\ln\left(\frac{\sinh \gamma_{13}}{\sinh \psi}\right)\right]\,\psi\,.
\end{eqnarray}
Note that $I_{cth-1}$ is exponentially suppressed in $\psi$ 
when $\psi\sim\gamma_{13}$.  However, for small $\psi$ the 
factor 
$\frac{\sinh 2\gamma_{13}}{\sinh ^2 \gamma_{13}-\sinh ^2 \psi}
=2+\mathcal{O}(e^{-\gamma_{13}})$. 
Therefore we can rewrite $I_{cth-1}$ as
\begin{eqnarray}
I_{cth-1}&=&\int_0^{\infty}d\psi\left[2\,\ln\left(\frac{2\sinh \gamma_{13}}{2\sinh \psi}\right)\right]\,\psi\,(\coth \psi-1)+\mathcal{O}(e^{-\gamma_{13}})\nonumber\\
&=&2\Biggl[ \gamma_{13}\int_0^{\infty}d\psi\,\psi\,(\coth \psi-1)\nonumber\\
&\ &\hspace{4mm}-\int_0^{\infty}d\psi\,\psi^2(\coth\psi-1)\nonumber\\
&\ &\hspace{4mm}-\int_0^{\infty}d\psi
 \,\psi\,\ln(1-e^{-2\psi})\,(\coth\psi-1) \Biggr]\nonumber\\
&\equiv& 2 ( k_1\gamma_{13} + k_2 + k_3)\,.
\end{eqnarray}
We evaluate $k_1$, $k_2$ and $k_3$ separately, starting with
\begin{eqnarray}
k_1&=&\int_0^{\infty}d\psi\,\psi \, (\coth\psi-1)\nonumber\\
&=&2\int_0^{\infty}d\psi\,\psi\frac{e^{-2\psi}}{1-e^{-2\psi}}\nonumber\\
&=&2\sum_{n=0}^{\infty}\int_0^{\infty}d\psi\,\psi\,e^{-2(n+1)\psi}\nonumber\\
&=&\frac{1}{2}\zeta(2)\,.
\end{eqnarray}
We evaluate $k_2$ and $k_3$ by using the same expansion with answers
\begin{eqnarray}
k_2&=&-\int_0^{\infty}d\psi\,\psi^2(\coth\psi-1)\nonumber\\
&=&-\frac{1}{2}\zeta(3)\,,
\end{eqnarray}
and finally
\begin{eqnarray}
k_3&=&-\int_0^{\infty}d\psi\,\psi\,\ln(1-e^{-2\psi})\,(\coth\psi-1)\nonumber\\
&=&\frac{1}{2}\zeta(3) \,.
\end{eqnarray}
Combining these results one finds
\begin{equation}
\label{Eq:cth-1}
I_{cth-1}=\zeta(2) \gamma_{13} \,.
\end{equation}

Now let's look at the remaining integral in 
Eq.~(\ref{Eq:I2_rewritten}), $I_1$. 
After some trivial algebra one can rewrite $I_1$ as
\begin{equation}
I_1=k_4+k_5\,,
\end{equation}
where
\begin{eqnarray}
k_4&=&2\int_0^{\gamma_{13}}d\psi\,\psi(\gamma_{13}-\psi)\,\frac{1}{(1-e^{-(\gamma_{13}-\psi)})(1+e^{-(\gamma_{13}-\psi)})}+\mathcal{O}(e^{-2\gamma_{13}})\nonumber\\
&=&2\gamma_{13}\int_0^{\gamma_{13}}d\lambda
\,\lambda\sum_{n=0}^{\infty}e^{-2n\lambda}
 - 2\int_0^{\gamma_{13}}d\lambda
\,\lambda^2\sum_{n=0}^{\infty}e^{-2n\lambda}
+\mathcal{O}(e^{-2\gamma_{13}})\,,
\end{eqnarray}
with $\lambda\equiv\gamma_{13}-\psi$, from which
\begin{equation}
k_4=2\left[\frac{\gamma_{13}^3}{6}+\gamma_{13}\frac{\zeta(2)}{4}-\frac{\zeta(3)}{4}\right]\,.
\end{equation}
Finally $k_5$ is given by
\begin{eqnarray}
k_5&=&-2\int_0^{\infty}d\psi\,\psi\,\ln(1-e^{-2\psi})\nonumber\\
&=&\frac{\zeta(3)}{2}\,,
\end{eqnarray}
by using same kind of manipulations. Combining the above results one finds
\begin{equation}
\label{Eq:remaining}
I_1=\frac{\gamma_{13}^3}{3} 
+ \frac{\zeta(2)}{2} \gamma_{13} \,.
\end{equation}
Using Eqs.~(\ref{Eq:cth-1}) and (\ref{Eq:remaining}), one finds
for the asymptotic behavior of $I_2$
\begin{equation}
\label{Eq:I2}
I_2(v_1,v_3)= \frac{\gamma_{13}^3}{3}
+ \frac{3\zeta(2)}{2} \gamma_{13}
+\mathcal{O}(e^{-\gamma_{13}})\,.
\end{equation}
By using this result in the expression for the single-pole term,
$W_{2E,3g-D}^{s.p.}$, we find
\begin{eqnarray}
\label{Eq:W2e3g_d}
W_{2E,3g-D}^{s.p.} &=& \left(\frac{\alpha_s}{\pi}\right)^2
d^{[t]}_{JI} \, C_A \,\frac{1}{2}\Bigg\{ - \frac{1}{4\vep} \, \gamma_{13}
+\frac{1}{16\vep}\left[ \frac{2}{3}\gamma_{13}^3
                      + 4\zeta(2)\gamma_{13} \right]\Bigg\}
+\mathcal{O}(e^{-\gamma_{13}})
\nonumber\\
&=&  - \left(\frac{\alpha_s}{\pi}  \right)^2
\, d^{[t]}_{JI} \, \frac{C_A}{2}\,
 \frac{1}{4\vep} \Biggl[ 
- \frac{\gamma_{13}^3}{6} 
+ \left(1 - \zeta(2)\right) \gamma_{13} \Biggr]
+ \mathcal{O}(e^{-\gamma_{13}})\,.
\end{eqnarray}
The contribution of the $\bar V_{\mu\nu\rho}$ piece to 
the diagram in Fig.~\ref{Fi:2ld}$(f)$ is given by~\cite{KR87}
\begin{equation}
\label{Eq:W2e3g_vbar}
W_{2E,3g-\bar V}^{s.p.}=-\left(\frac{\alpha_s}{\pi}\right)^2
\, d^{[t]}_{JI} \,\frac{C_A}{2}\,\frac{1}{4\vep}
\,\left[ - \gamma_{13} + \frac{\zeta(2)}{4}
+ \frac{1}{2}\,I_3(\gamma_{13})
+ \mathcal{O}(e^{-\gamma_{13}})\right]\,,
\end{equation}
where
\begin{equation}
I_3(\gamma_{13})\equiv \sinh (2\gamma_{13})\,\int_0^{\gamma_{13}}d\psi\,\frac{1}{\sinh^2\gamma_{13}-\sinh^2\psi}\ln\left(\frac{\sinh \gamma_{13}}{\sinh \psi}\right)\,.
\end{equation}
One can analyze the high-energy asymptotics of $I_3$ in a similar 
way as above, with the result
\begin{equation}
\label{Eq:I3}
I_3(\gamma_{13}) = \gamma_{13}^2
+ \frac{3\zeta(2)}{2} + \mathcal{O}\left(e^{-\gamma_{13}}\right)\,.
\end{equation}
Combining Eqs.~(\ref{Eq:W2e3g_d}),~(\ref{Eq:W2e3g_vbar}) 
and (\ref{Eq:I3}), and letting $\gamma_{13} = T$, we find
\begin{eqnarray}
W_{2E,3g}^{s.p.}
&=&  - \left(\frac{\alpha_s}{\pi}  \right)^2
\, d^{[t]}_{JI} \, \frac{C_A}{2}\,
 \frac{1}{4\vep}
\Bigg\{\, \left[ 
- \frac{T^3}{6} + \left(1 -  \zeta(2) \right) T
\right]
\nonumber\\
&\ & \hspace{40mm} + \left[\frac{T^2}{2}-T+\zeta(2) \right]\, 
\Bigg \}+\mathcal{O}\left(e^{-\gamma_{13}}\right)\,,
\label{Ff2loopsAPP}
\end{eqnarray}
which is the result given in Eq.~(\ref{Ff2loops}).

The amplitudes for diagrams $(d)$ and $(e)$ of Fig.~\ref{Fi:2ld} are 
given in Ref.~\cite{KK95} and the high-energy asymptotics is obtained with 
a similar analysis. The results are given in Eq.~(\ref{Fde2loops}).
As mentioned above, these contributions cancel the 
$\bar V_{\mu\nu\rho}$ contribution from diagram $(f)$,
which is enclosed by the second set of brackets in Eq.~(\ref{Ff2loopsAPP}).

Finally, let us look at the crossed-ladder diagram in 
Fig.~\ref{Fi:2ld}$(b)$. The velocity factor in configuration 
space is given by
\begin{eqnarray}
F_{CL,t}(v_1,v_3)=(ig\mu^{\vep})^4\,(v_1\cdot v_3)^2\int_0^{\infty}d\alpha_1\int_0^{\alpha_1}d\alpha_2\int_0^{\infty}d\beta_1\int_0^{\beta_1}d\beta_2\nonumber\\
&\ &\hspace{-70mm}\times D(v_1\alpha_1+v_3\beta_2)D(v_1\alpha_2+v_3\beta_1)\,.
\label{FCLt}
\end{eqnarray}
It is not difficult to show that
the single-pole part of the crossed-ladder velocity factor 
is precisely the negative of that for the uncrossed-ladder diagram
in Fig.~\ref{Fi:2ld}$(a)$.  Therefore 
in the combination of the two diagrams
the single-pole part of Eq.~(\ref{FCLt}) is 
multiplied by the difference of the respective color factors.
Although the individual color factors are not proportional to the
one-loop factor $d^{[t]}_{JI}$, their difference evaluates to 
$d^{[t]}_{JI} C_A/2$.   The following result for the combination
of diagrams $(a)$ and $(b)$ can also be 
found in Ref.~\cite{KR87},\footnote{%
Needless to say we can evaluate the integrals in momentum space 
and get the same result.}
\begin{equation}
\label{Eq:CLamp}
W_{CL+L,t}^{s.p.}=-\left(\frac{\alpha_s}{\pi}\right)^2
\,d^{[t]}_{JI} \frac{C_A}{2}
\,\frac{1}{2\vep} \, \coth ^2\gamma_{13} \, I_3(v_1,v_3)\,,
\end{equation}
where we define
\begin{equation}
I_3(v_1,v_3)\equiv\int_0^{\gamma_{13}}d\psi
\,\psi\,(\gamma_{13}-\psi)\,\coth\psi\,.
\end{equation}
One can investigate the asymptotic behavior of $I_3$ in a
way similar to that presented for the $3$-gluon vertex
diagram in Fig.~\ref{Fi:2ld}$(f)$.  One obtains the result
\begin{equation}
I_3(v_1,v_3) = \frac{\gamma_{13}^3}{6}
+\frac{\zeta(2)}{2} \gamma_{13}
- \frac{\zeta(3)}{2}
+\mathcal{O}(e^{-\gamma_{13}})\,.
\end{equation}
By using the above relation in Eq.~(\ref{Eq:CLamp}) we find
\begin{equation}
W_{CL+L,t}^{s.p.}=-\left(\frac{\alpha_s}{\pi}\right)^2
\, d^{[t]}_{JI} \frac{C_A}{2}
\, \frac{1}{2\vep}\left( 
 \frac{\gamma_{13}^3}{6} 
 + \frac{\zeta(2)}{2} \gamma_{13}
 - \frac{\zeta(3)}{2}\right)+\mathcal{O}(e^{-\gamma_{13}})\,.
\end{equation}
Letting $\gamma_{13} = T$, this is the result given in 
Eq.~(\ref{Fabc2loops}), along with the result for 
diagram $(c)$~\cite{KR87}.

\section{The commutator of ${\bf I}_{\f}^{(1){\rm fin}}$ 
and ${\bf\Gamma}_{S_\f}^{(1)}$}
\label{comapp}

The task of this appendix is to evaluate the commutator 
$\left [ \I^{(1){\rm fin}}_\f \, ,\, {\bf\Gamma}_{S_\f}^{(1)} \right ]$
appearing on the left-hand side of Eq.~(\ref{commutator}).
Note that the pole parts of $\I_{\f}^{(1)}$
can be identified with ${\bf\Gamma}_{S_\f}^{(1)}$, via Eq.~(\ref{I1expand}).
Writing out the ${\cal O}(\vep^0)$ parts of $\I_{\f}^{(1){\rm fin}}$ with
nontrivial color structure, the commutator of the finite and pole parts of 
$\I_{\f}^{(1)}$ becomes,
\bea
&& \left [   \I^{(1){\rm fin}}_\f \, ,\, {\bf\Gamma}_{S_\f}^{(1)}    \right ]
=\nonumber\\
&& \quad  \frac{1}{4}\,
\Bigg [\,  \sum_k
\sum_{l\ne k}({\bf T}_k\cdot {\bf T}_l)\, \left( \frac{1}{2}\ln^2 \left(\frac{\mu^2}{-s_{kl}}\right)
+ \frac{\gamma_k}{{\bf T}^2_k}\, \ln \left(\frac{\mu^2}{-s_{kl}}\right)\right)
\, ,\, 
\sum_i\sum_{j\ne i}({\bf T}_i\cdot {\bf T}_j)
  \,\ln\left(\frac{\mu^2}{-s_{ij}}\right) \, \Bigg ]
  \nonumber\\
  && = \ {\cal C}_{3,\f}\  + \ {\cal C}_{2,\f}\, ,
 \label{commutator2}
\eea
where in the second equality we introduce notation to separate the
terms with three logarithms (${\cal C}_{3,\f}$) from those with
two (${\cal C}_{2,\f}$).   

In the case where all external lines are gluons, or all
are quarks and/or antiquarks, all the 
ratios ${\gamma_k}/{{\bf T}^2_k}$ in the left-hand side
of the commutator are equal.  This term is then proportional
to ${\bf\Gamma}_{S_\f}^{(1)}$, and ${\cal C}_{2,\f}$ vanishes.
This argument does not apply, of course, to mixed processes,
such as $q\bar{q}\to gg$.  For the latter case, however, and
for any other $2\to 2$ process, we may use the color
conservation identity $\sum_k {\bf T}_k=0$ and the
simplicity of the kinematics to show that
${\cal C}_{2,\f}$  vanishes.   The argument is
simple, and may be given for the case $q\bar{q}\to gg$
without loss of generality.  In this case, we may take
$k=1,2$ in Eq.~(\ref{commutator2}) to correspond to the 
incoming quark and antiquark,
and we consider just these terms in the double-logarithmic
part of the commutator in Eq.~(\ref{commutator2}).   
We focus first on the terms with prefactor 
${\gamma_q}/{{\bf T}^2_q}=3/2$.  (The same argument applies
to the remaining terms, with prefactor ${\gamma_g}/{{\bf T}^2_g}$.)
These terms are proportional to
\bea
\Bigg [\,  \sum_{k=1}^2 \sum_{l\ne k}
{\bf T}_k\cdot {\bf T}_l\, 
  \ln \left(\frac{\mu^2}{-s_{kl}}\right)
\, ,\, 
\sum_{i=1}^4\sum_{j\ne i}\, {\bf T}_i\cdot {\bf T}_j
  \,\ln\left(\frac{\mu^2}{-s_{ij}}\right) \, \Bigg ]\, .
\label{commutator3}
\eea
This commutator would vanish if the sum over index $k$
were extended to $k=3$ and 4.  But this can by done
by observing that $\sum_k {\bf T}_k=0$ implies that,
for example,
\bea
{\bf T}_1\cdot {\bf T}_2 = {\bf T}_3\cdot {\bf T}_4+\frac{1}{2}\left(T_3^2+T_4^2-T_1^2-T_2^2\right)\, ,
\eea
where the squared terms commute with all combinations of generators.  At the
same time, we have $s_{12}=s_{34}$.  As a result, in the
left-hand term of the commutator~(\ref{commutator3}), we may make 
the replacement
\bea
{\bf T}_1\cdot {\bf T}_2\, 
\ln \left(\frac{\mu^2}{-s_{12}}\right) \to {\bf T}_3\cdot {\bf T}_4\,
  \ln \left(\frac{\mu^2}{-s_{34}}\right)\, .
\eea
Analogous reasoning for each of the terms in the
sum over $k$ and $l$ in Eq.~(\ref{commutator3}) shows
that for $2\to 2$ scattering the sum of the
missing terms with $k=3,4$ is identical in the commutator to
the sum from $k=1,2$.   Inserting the missing terms, at the price
of an overall factor of $1/2$, the two entries of the commutator 
become identical and it vanishes.
This argument, of course, is heavily dependent on the
specifics of $2\to 2$ scattering.   We know of no
general argument that would eliminate all double-logarithmic
terms in the commutator in $2\to n$ processes; indeed
such terms are generically present.

We now consider the triple-logarithmic terms in Eq.~(\ref{commutator2}),
\bea
 {\cal C}_{3,\f} 
&=&  \frac{1}{8}\,
\Bigg [\,  \sum_k
\sum_{l\ne k}\, {\bf T}_k\cdot {\bf T}_l\, ,\, 
\sum_i\sum_{j\ne i}\, {\bf T}_i\cdot {\bf T}_j
 \, \Bigg ]\, b_{kl}\, a_{ij}
  \nonumber\\
&=& 
 \frac{1}{2}\, \sum_{i\ne j\ne k}\Big[\, 
{\bf T}_k\cdot {\bf T}_j\, ,\, 
{\bf T}_j\cdot {\bf T}_i
 \, \Big]\, b_{kj}\, a_{ij}
\nonumber\\
&=&
 \frac{1}{2}\, \sum_{i\ne j\ne k}   \,
 i\, f_{a_1a_2a_3}\,  
{\bf T}_k^{a_1}\,  {\bf T}_j^{a_3}\, 
 {\bf T}_i^{a_2}
\, b_{kj}\, a_{ij}\, ,
 \label{commutatorf}
\eea
where in the first equality we have introduced the notation
$b_{kl} = \ln^2(\mu^2/(-s_{kl})) = b_{lk}$ 
and $a_{ij} = \ln(\mu^2/(-s_{ij})) = a_{ji}$.
In the second equality in Eq.~(\ref{commutatorf}) 
we have identified the nonvanishing
terms in the commutator, for which one and only one
pair of generators is matched between the two entries
of the commutator.  Because the scalar products
are symmetric, there are four ways in which this
matching may occur, for fixed indices $i\ne j\ne k$.
Finally, the third equality shows the result of
performing the commutator explicitly for the
generators on the $j$ line.  This form is reminiscent
of the color structure of $\hat{\bf H}_\f^{(2)}$, Eq.~(\ref{H2def}), although
the triple-logarithmic momentum factors are different, and depend
on the renormalization scale, $\mu$.

To make contact between Eq.~(\ref{commutatorf}) and
the explicit expression~(\ref{H2def}) for $\hat{\bf H}_\f^{(2)}$, 
we convert the sum of
unequal choices of $i,\, j$ and $k$ into a sum
over distinguishable triplets, denoted $(i,j,k)$.  
For each such choice, there are six permutations
of the indices $i,\, j$ and $k$ in the final expression
of Eq.~(\ref{commutatorf}).  These can be
thought of as three cyclic permutations,
which leave the structure constants the same,
but change the momentum factors, and
three more (exchanges of $i$ and $k$ for fixed
$j$, plus cyclic permutations), which change the sign of the 
structure constants, and change the kinematic factors.

Following this path, we define
\bea
c_{[kj,ji]} \equiv b_{kj}\, a_{ji} \ -\ b_{ij}\, a_{jk} 
\eea
and rewrite 
 ${\cal C}_{3,\f}$ as
 \bea
 {\cal C}_{3,\f} 
&=& 
 \frac{1}{2}\, \sum_{(i, j, k)}   \,
 i\, f_{a_1a_2a_3}\,  
{\bf T}_k^{a_1}\,  {\bf T}_j^{a_3}\, 
 {\bf T}_i^{a_2}
\, \Big[\, c_{[kj,ji]} + c_{[ik,kj]} + c_{[ji,ik]}\, \Big] \, .
 \label{commutatorc}
\eea
A straightforward calculation shows that all
of the $\mu$-dependence cancels
in this expression.  Relabelling the indices, and using the
antisymmetry of the structure constants, we derive
\bea
 {\cal C}_{3,\f} 
&=&
- \frac{i}{2}\, \sum_{(i, j, k)}\, f_{a_1a_2a_3}\,  
{\bf T}_i^{a_1}\,  {\bf T}_j^{a_2}\, 
 {\bf T}_k^{a_3}\, 
\ln\left(\frac{-s_{ij}}{-s_{jk}}\right)\,
\ln\left(\frac{-s_{jk}}{-s_{ki}}\right)\,
\ln\left(\frac{-s_{ki}}{-s_{ij}}\right)
\nonumber\\
&=& -\, \frac{1}{2}\, \hat{\bf H}_\f^{(2)} \,,
\eea
which establishes the result of Eq.~(\ref{commutator}).  We emphasize
that, unlike our demonstration that ${\cal C}_{2,\f}$ vanishes,
this result holds for an arbitrary $2\to n$ process.

\end{appendix}

\end{document}